\documentclass[prl,aps,twocolumn,superscriptaddress,showpacs,footinbib,longbibliography]{revtex4-2}
\usepackage[colorlinks=true,linkcolor=black, citecolor=blue, urlcolor=blue, 
    unicode=true]{hyperref}

\usepackage{graphicx}
\usepackage{epstopdf}
\usepackage{color}

\begin{document}

\title{Magnetoelastic dynamics of the ``spin Jahn-Teller'' transition in CoTi$_{2}$O$_{5}$}

\author{K. Guratinder}
\affiliation{School of Physics and Astronomy, University of Edinburgh, Edinburgh EH9 3JZ, UK}
\author{R. D. Johnson}
\affiliation{Department of Physics and Astronomy, University College London, Gower St., London, WC1E 6BT, UK}
\affiliation{London Centre for Nanotechnology, University College London, Gordon Street, London WC1H 0AH, UK}
\author{D. Prabhakaran}
\affiliation{Department of Physics, Clarendon Laboratory, University of Oxford, Parks Road, Oxford OX1 3PU, UK}
\author{R. A. Taylor}
\affiliation{Department of Physics, Clarendon Laboratory, University of Oxford, Parks Road, Oxford OX1 3PU, UK}
\author{F. Lang}
\affiliation{Department of Physics, Clarendon Laboratory, University of Oxford, Parks Road, Oxford OX1 3PU, UK}
\author{S. J. Blundell}
\affiliation{Department of Physics, Clarendon Laboratory, University of Oxford, Parks Road, Oxford OX1 3PU, UK}
\author{L. S. Taran}
\affiliation{M. N. Mikheev Institute of Metal Physics, Ural Branch of Russian Academy of Sciences, 620137 Yekaterinburg, Russia}
\author{S. V. Streltsov}
\affiliation{M. N. Mikheev Institute of Metal Physics, Ural Branch of Russian Academy of Sciences, 620137 Yekaterinburg, Russia}
\affiliation{Institute of Physics and Technology, Ural Federal University, 620002 Yekaterinburg, Russia}
\author{T. J. Williams}
\affiliation{ISIS Pulsed Neutron and Muon Source, STFC Rutherford Appleton Laboratory, Harwell Campus, Didcot, Oxon OX11 0QX, United Kingdom}
\author{S. R. Giblin}
\affiliation{School of Physics and Astronomy, Cardiff University, Cardiff CF24 3AA, UK}
\author{T. Fennell}
\affiliation{Laboratory for Neutron Scattering, Paul Scherrer Institut, CH-5232 Villigen, Switzerland}
\author{K. Schmalzl}
\affiliation{Julich Centre for Neutron Science, Forschungszentrum Julich GmbH, Outstation at Institut Laue-Langevin, Boite Postal 156, 38042 Grenoble Cedex 9, France}
\author{C. Stock}
\affiliation{School of Physics and Astronomy, University of Edinburgh, Edinburgh EH9 3JZ, UK}

\date{\today}

\begin{abstract}
	
	CoTi$_{2}$O$_{5}$ has the paradox that low temperature static magnetic order is incompatible with the crystal structure owing to a mirror plane that exactly frustrates magnetic interactions.  Despite no observable structural distortion with diffraction, CoTi$_{2}$O$_{5}$ does magnetically order below $T_{\rm N}$ $\sim$ 25 K with the breaking of spin ground state degeneracy proposed to be a realization of the spin Jahn-Teller effect in analogy to the celebrated orbital Jahn-Teller transition.  We apply neutron and Raman spectroscopy to study the dynamics of this transition in CoTi$_{2}$O$_{5}$.  We find anomalous acoustics associated with a symmetry breaking strain that characterizes the spin Jahn-Teller transition. Crucially, the energy of this phonon coincides with the energy scale of the magnetic excitations, and has the same symmetry of an optic mode, observed with Raman spectroscopy, which atypically softens in energy with decreasing temperature. Taken together, we propose that the energetics of the spin Jahn-Teller effect in CoTi$_{2}$O$_{5}$ are related to cooperative magnetoelastic fluctuations as opposed to conventional soft critical dynamics which typically drive large measurable static displacements.
	
\end{abstract}


\maketitle

Structural phase transitions are typically driven by a dynamic soft mode~\cite{Cochran60:9,Cochran61:10,Shirane74:46,Cowley06:75} which freezes either continuously or discontinuously.  Classic examples include soft zone center transverse optical phonons in perovskites~\cite{Shirane70:2,Harada70:608,Harada71:4} and zone boundary phonon anomalies that appear in orientational-like order-disorder transitions~\cite{Yamada74:9,Harris98:10}.  Such structural transitions are readily observed through diffraction techniques.  However, there is a growing list of materials where a magnetostructural transition must occur based on symmetry constraints or bulk measurements, yet are not accompanied by an observable change in the Bragg peaks measured with X-ray or neutron diffraction.  For example, this is the case for ferroaxial multiferroic phase transitions in RbFe(MoO$_{4}$)$_2$~\cite{Kenzelmann07:98,Hearmon12:108},  CaMn$_{7}$O$_{12}$~\cite{Johnson12:108,Kadlec14:90}, and Cu$_{3}$Nb$_{2}$O$_{8}$~\cite{Johnson11:107,Giles-Donovan20:102} where changes in Bragg peaks and soft lattice dynamics are not readily observable~\cite{Kambda21:9}.  This is despite known low temperature structural transitions derived based on symmetry constraints.  A further example is certain ferroelectric metal-organic frameworks~\cite{Ding23:7} which lack observable soft phonon dynamics despite known displacive ferroelectric transitions. 

We address this apparent contradiction through studying the lattice and magnetic dynamics of a similar problem in the spin-Jahn Teller material CoTi$_{2}$O$_{5}$ where a structural transition is not observable in conventional diffraction, yet must occur to allow low temperature magnetic order which is exactly frustrated based on the high temperature nuclear structure. In analogy to the orbital Jahn-Teller theorem~\cite{Gehring75:38} which predicts a structural distortion will occur to break an orbital degeneracy, a similar idea has been proposed in the context of spin degeneracy from frustrated magnetism where an orbital degeneracy is absent. The ``spin Jahn-Teller" effect is where a spin-degeneracy drives a symmetry breaking structural transition that lifts the degeneracy of the spin-manifold and facilitates long-range magnetic order.  Such a structural distortion originates from the competition between magnetic and elastic energy scales.~\cite{Tchernyshyov02:88,Yamashita00:85}  Given the reduction of the magnetic energy scales linearly with atomic positions while the elastic scales quadratically, it is favored for distortion to occur for the relief of frustrating magnetic interactions, resulting in spatially long-range magnetic order.  We study the dynamics of this effect in CoTi$_{2}$O$_{5}$ by applying neutron and Raman spectroscopy.  While a structural transition is not observable, we find evidence of acoustic dynamics associated with atomic displacement that is linked with magnetic frustration.

\begin{figure*}[t]
	\begin{center}
		\includegraphics[width=155mm,trim=0.2cm 0.0cm 0.0cm 0cm,clip=true]{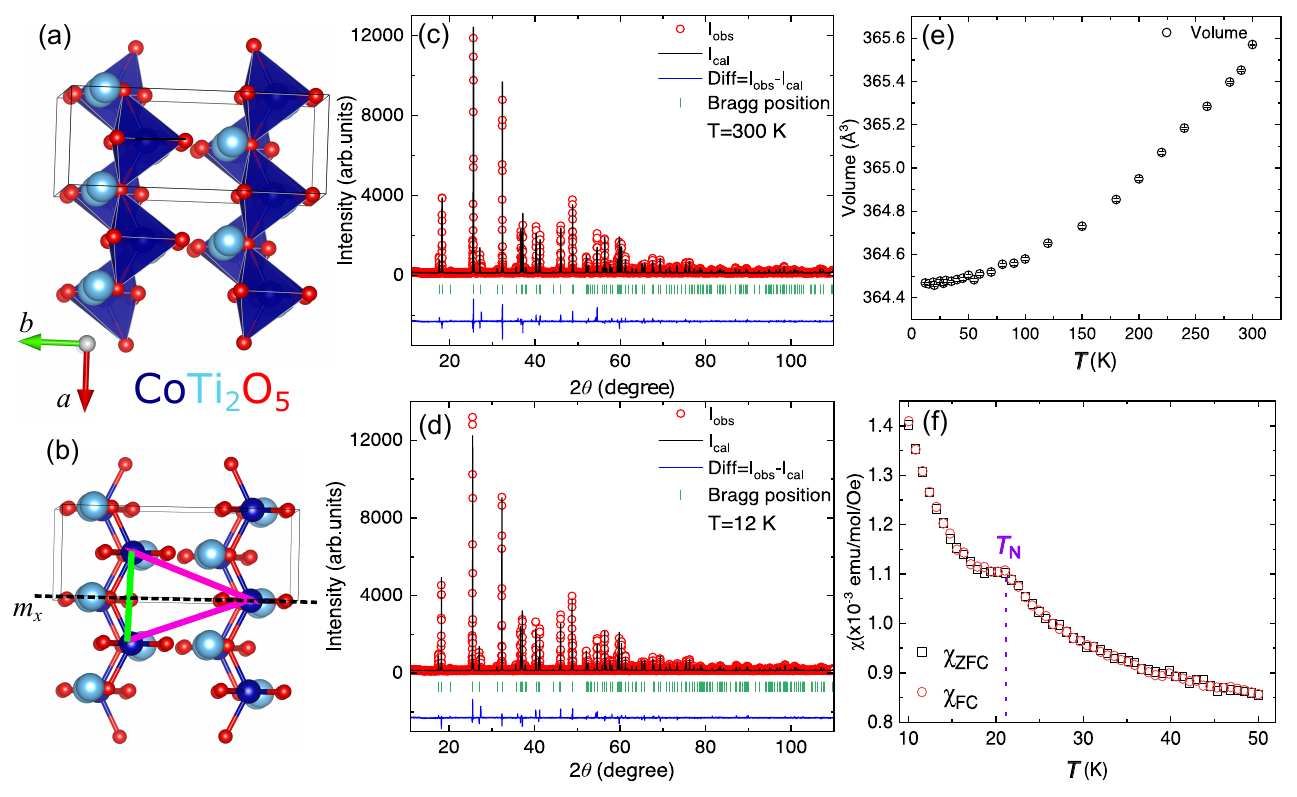}
	\end{center}
	\caption{Structural and magnetic properties of CoTi$_{2}$O$_{5}$. $(a,b)$ illustrates the nuclear structure and the frustrating triangular arrangement of the interactions.  $(c,d)$ illustrate the same diffraction measurements at 300 and 12 K. $(e)$ illustrates a plot of the volume of the lattice with temperature showing no change in slope at the $(f)$ N\'eel temperature of $\sim$ 23 K measured in a powder sample with an applied field of 100 Gauss.} 
	\label{fig:structural}
\end{figure*}

CoTi$_{2}$O$_{5}$ adopts a pseudobrookite structure with an orthorhombic space group $Cmcm$ at high temperatures.~\cite{Muller83:114}  The magnetic Co$^{2+}$ ions reside on a site with $m2m$ ($C_{2\nu}$) symmetry, implying that the orbital levels are already nondegenerate and therefore the normal Jahn-Teller theorem~\cite{Streltsov20:10} discussed above does not apply.  The lack of an orbital degeneracy in CoTi$_{2}$O$_{5}$ is further discussed in the Supplementary Information (SI) in the context of the search for spin-orbit transitions seen in octahedrally coordinated and orbitally active Co$^{2+}$ based compounds~\cite{Cowley13:88,Wallington15:92,Sarte18:98,Sarte18:98_2,Songvilay20:102} which we find are absent in CoTi$_{2}$O$_{5}$.  The frustrated antiferromagnetic interactions, mediated by indirect exchange through oxygen, are illustrated in Fig. \ref{fig:structural} $(a,b)$ and are based on perfectly isosceles triangles.  Because of a crystallographic mirror plane perpendicular to [100] (termed $m_{x}$), this is a perfectly frustrated geometry resulting in spin order that is expected to be unstable to thermal fluctuations.  Therefore, magnetic order is not expected unless accompanied by a structural distortion (displacive and strain)~\cite{Stock09:103} that transforms as the $\Gamma_{2}^{+}$ irreducible representation as identified in Ref. \onlinecite{Kirschner19:99}.  However, as shown in Ref. \onlinecite{Kirschner19:99} applying high resolution neutron diffraction, CoTi$_{2}$O$_{5}$ does not undergo an observable structural transition yet displays spatially long-range magnetic order below $T_{\rm N}$ $\sim$ 25 K.  It was therefore concluded that CoTi$_{2}$O$_{5}$ undergoes a spin Jahn-Teller transition where the structure distorts, subtly, to break the large ground state spin degeneracy imposed by the frustrating geometry.  Further studies have been performed on isostructural FeTi$_{2}$O$_{5}$ and have also found similar results.~\cite{Lang19:100}  In the following, we investigate acoustic and optic structural dynamics near this spin Jahn-Teller transition and probe low-energy magnetic excitations that characterize low temperature magnetic order.  

\begin{figure}[t]
	\begin{center}
		\includegraphics[width=80mm,trim=3.0cm 4.25cm 3cm 4.5cm,clip=true]{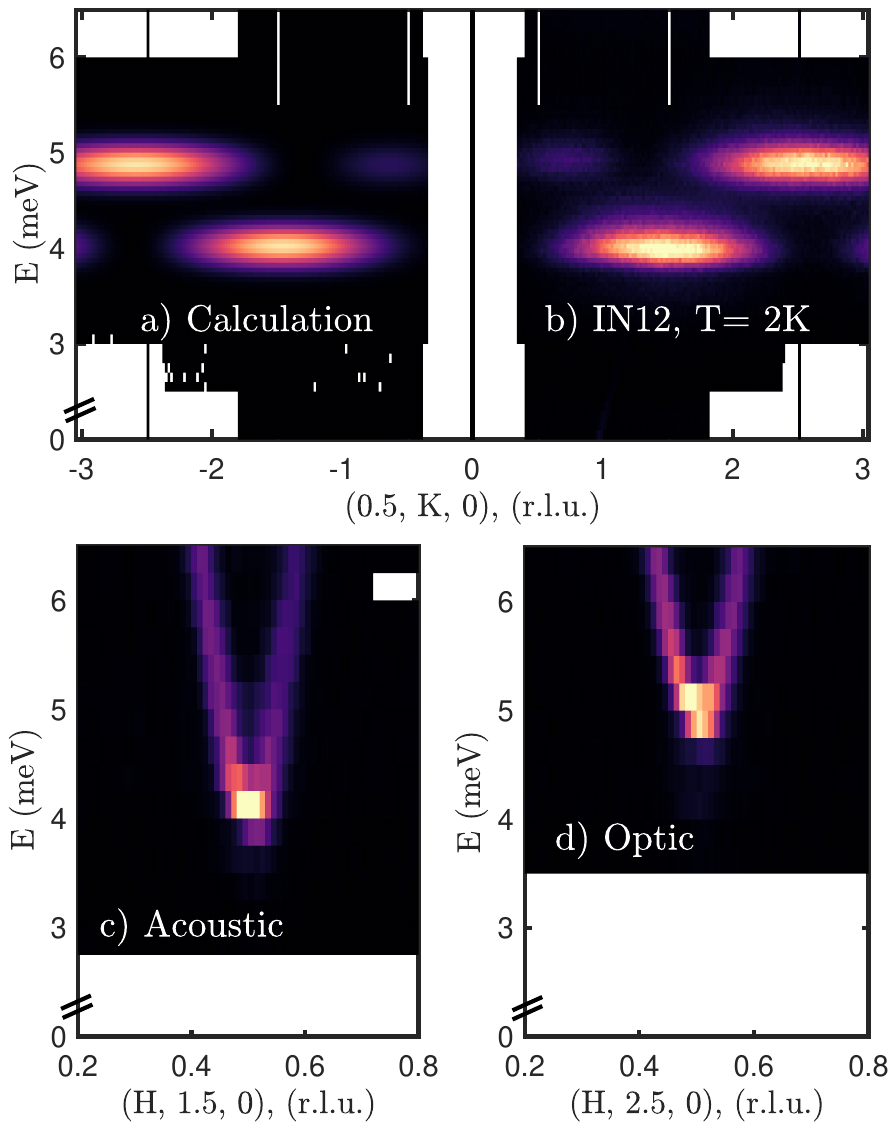}
	\end{center}
	\caption{The low-energy magnetic excitations in the magnetically ordered phase. Given there are two magnetic ions in the unit, there are two spin-wave branches which disperse along (H,0,L) with differing phase, termed acoustic and optic here. $(a)$ Calculated structure factors and $(b)$ measured magnetic correlations perpendicular to the chain axis illustrating non-measurable dispersion indicating weak coupling.  This contrasts with the resulting coupling along the chain direction displayed in $(c,d)$ for each of the two magnetic domains.} 
	\label{fig:magnetic_dynamics}
\end{figure}

Neutron spectroscopy was used to study the low-energy acoustic phonons and magnetic excitations.  Acoustic phonon measurements were performed on the EIGER thermal triple-axis spectrometer (PSI).  High resolution magnetic spectroscopy was performed on the IN12 cold-triple axis spectrometer (ILL).  For all neutron scattering experiments, the sample was aligned such that reflections of the form (H,K,0) were within the horizontal plane.  We discuss the magnetic dispersion along the (0,0,L) direction in the SI applying the time of flight MAPS spectrometer. The sample was grown using the traveling floating zone technique \cite{Kirschner19:99}.  Monochromatic X-ray diffraction as a function of temperature was carried out on a Rigaku Smartlab with a Johansson monochromator and PheniX displex.  Temperature dependent Raman spectroscopy was performed to study optical phonons. Further experimental information is in the SI.

\begin{figure}[t]
	\begin{center}
		\includegraphics[width=83mm,trim=3.3cm 4.5cm 4.0cm 4.0cm,clip=true]{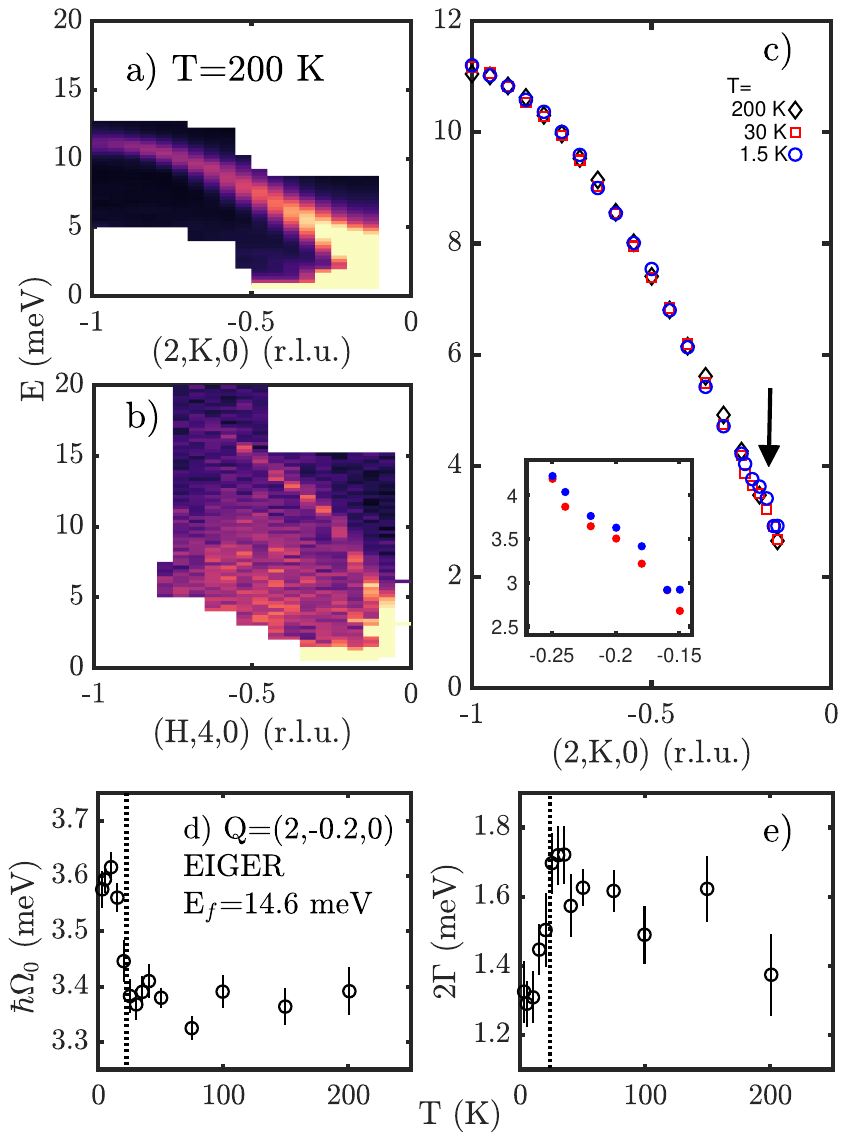}
	\end{center}
	\caption{Acoustic phonons in CoTi$_{2}$O$_{5}$ with the dispersion at T=200 K for acoustic phonons propagating along $b^{*}$ $(a)$ and $a^{*}$ $(b)$ (note the momentum and energy broadened paramagnetic scattering near $H\sim$-0.5). An anomaly in the dispersion is evident at $\vec{Q}\sim$(2, -0.2, 0) $(c)$. $(d)$ and $(e)$ illustrate the energy position ($\hbar \Omega_{0}$) and full-width at half maximum ($2\Gamma$) as a function of temperature.  } 
	\label{fig:phonons}
\end{figure}

Figure \ref{fig:structural} reviews the static structural and magnetic properties of CoTi$_{2}$O$_{5}$.  The nuclear structure summarized in Figs. \ref{fig:structural} $(a-b)$ is built on chains of Co$^{2+}$ ions along the crystallographic $a$ axis.  The chains are coupled along the crystallographic $b$ axis via an isosceles triangular arrangement which is exactly frustrated.  Monochromatic X-ray diffraction at 300 K (\ref{fig:structural} $c$) is compared to 12 K (Fig. \ref{fig:structural} $d$) with no structural distortion observed between these two temperatures (refinement and analysis discussed in SI~\cite{Petricek14:229,TOPAS}).  This is further confirmed in Fig. \ref{fig:structural} $(e)$ where we plot the unit cell volume as a function of temperature.  At low temperatures, there is no measurable change in slope which if observed would indicate a structural distortion at $T_{\rm N}$.  However, as illustrated in Fig. \ref{fig:structural} $(f)$, despite the frustrating geometry of the spins, a magnetic transition occurs at $T_{\rm N}$ $\sim$ 23 K evidenced by a peak in the magnetic susceptibility.   

\begin{figure}[t]
	\begin{center}
		\includegraphics[width=85mm,trim=8.2cm 0.0cm 8.2cm 0cm,clip=true]{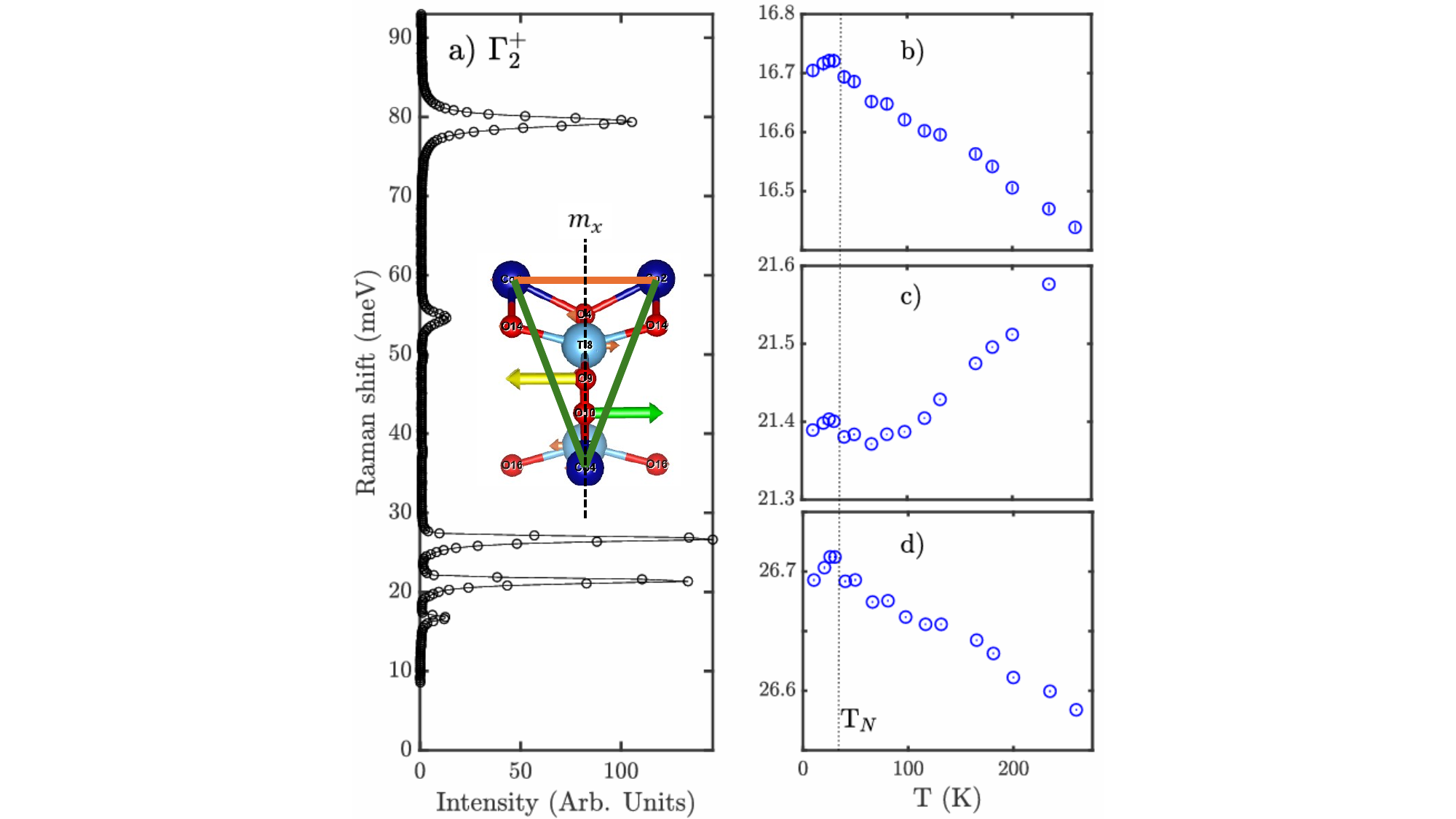}
	\end{center}
	\caption{$(a)$ Raman spectra sensitive to optic phonon modes with $\Gamma_{2}^{+}$ symmetry which break the frustrating $m_{x}$ mirror plane illustrated in the inset through the motion of the oxygen atoms.  $(b-c)$ illustrate temperature dependence of several modes with $(c)$ displaying a softening.} 
	\label{fig:raman}
\end{figure}

We now discuss the low-energy magnetic excitations below $T_{\rm N}$.  Energy-momentum slices along $K$ and $H$ are shown in Fig. \ref{fig:magnetic_dynamics} $(b)$ and $(c,d)$ respectively illustrating energetically gapped~\cite{Xu24:33} magnetic excitations. Confirming expectations based on bonding in the structure and the frustrating triangular arrangement outlined above, the excitations are dispersionless along the $K$ (Fig. \ref{fig:magnetic_dynamics} $b$) direction and strongly dispersive along $H$ (Fig. \ref{fig:magnetic_dynamics} $c,d$) where Co$^{2+}$ ions are arranged in one-dimensional chains with strong interactions along the crystallographic $a$ axis.  We note that the scans along $K$ display two distinct magnetic modes, as expected given the presence of two magnetic ions per primitive unit cell.   Fits to the low energy magnetic dispersion applying a nearest-neighbor anisotropic one-dimensional model give a coupling of 3.3$\pm$0.2 meV along the chain direction, an anisotropy $D$=0.25 $\pm$ 0.12 meV, and less than 0.07 meV along $b$ defined by the experimental resolution on IN12.   Further measurements discussed in the SI indicate a weak 0.3 $\pm$ 0.15 meV exchange along $c$.  The lack of any observable dispersion along the $K$ axis in Fig. \ref{fig:magnetic_dynamics} $(b)$ is consistent with no measurable magnetic exchange along the crystallographic $b$ axis.  This is expected given the frustrating geometry imposed by the $m_{x}$ mirror plane; however is unexpected given the presence of low-temperature spatially long-range magnetic order. 

To understand the intensity modulation along $K$ we compare the data to the nearest neighbor ``buckled sheet" model (SI - Fig. 5).  Considering two Co$^{2+}$ sites in a given $a-b$ plane, the neutron scattering intensity from optic and acoustic correlations would take the form of $I(K) \propto [1 \pm \cos(4\pi K y)]$, with $y\sim$0.1911.  The $\pm$ sign fixes the relative excitation phase for the two antiferromagnetic magnon modes in the $(H,L)$ plane with $-$ corresponding to acoustic-like (in-phase) magnetic fluctuations and $+$ being optic-like with out of phase-like fluctuations of the neighboring spins.  Fig. \ref{fig:magnetic_dynamics} $(a)$ is a calculation to this model fixing spin excitations at 4 and 4.85 meV.  Dispersive H slices for these two modes are shown in Fig. \ref{fig:magnetic_dynamics} $(c-d)$ (see SI for further details).


With a crystallographic distortion not observable with high resolution X-ray or neutron diffraction and no measurable dispersion of the spin excitations along $b$, we investigate the acoustic shear phonons in Fig. \ref{fig:phonons}.  Acoustic lattice fluctuations can be very sensitive to small structural distortions as motivated by recent studies linking low-energy acoustic phonons with weak nematic order in pnictides and chalcogenides~\cite{Wu21:126,Kauth20:102}, previous works on the dynamic Jahn-Teller effects in rare earth compounds~\cite{Lowenhaupt79:42,Guntherodt83:51,Thalmeier82:49}, and also the sensitivity of acoustic phonons to orbital Jahn-Teller effects~\cite{Melcher72:28,Pytte71:3,Pytte73:8,Liu19:122,Weber17:96,Lane23:5,Birgeneau74:10}.  Fig. \ref{fig:phonons} $(a)$ displays the shear mode with $\Gamma_{2}^{+}$ symmetry ($B_{1g}$ in the notation of Ref. \onlinecite{Cowley76:13}) and Fig. \ref{fig:phonons} $(b)$ the shear mode with $\Gamma_{4}^{+}$ symmetry ($B_{2g}$ in Ref. \onlinecite{Cowley76:13} notation).  The $\Gamma_{2}^{+}$ shear mode (Fig. \ref{fig:phonons} $a$) has a lower energy and is defined by a shear with unique crystallographic $c$ axis with the correct symmetry to break the $m_x$ mirror plan in Fig. \ref{fig:structural} $(b)$.  Given its lower energy scale and correct symmetry, we study the temperature dependence of this acoustic phonon in Fig. \ref{fig:phonons} $(c)$, noting a small anomaly in the dispersion at K$\sim$ -0.2 displayed in the inset. 

Figures \ref{fig:phonons} $(d)$ and $(e)$ plot the energy values ($\hbar\Omega_{0}$) and full-width ($2\Gamma$) against temperature obtained from constant $\vec{Q}$=(2, -0.2, 0) scans fit to a damped harmonic oscillator convolved with the experimental resolution (discussed in the SI). The measured N\'eel transition temperature, from susceptibility, is marked by the dashed line in both panels.  On cooling below $T_{\rm N}$, there is an abrupt hardening of the acoustic phonon at this position accompanied by a sharpening in energy indicative of increased phonon lifetime.  Given that in the limit $q\rightarrow 0$ the acoustic phonon velocity is related~\cite{Dove:book} to the elastic constant (in this case $C_{66}$~\cite{Cowley76:13}), the hardening of this acoustic phonon at low temperatures in the magnetically ordered phase is indicative of a dynamic strain.  We suggest this is a dynamic effect as no change on this scale (5-6 \%) is observed in the cell volume~\cite{Maradudin62:2} presented in Fig. \ref{fig:structural} or splitting of nuclear Bragg peaks on the scale reported for similar energy shifts in the acoustic phonons in other materials~\cite{Guratinder23:108}.

Having observed an anomaly in the acoustic phonons at $T_{\rm N}$ with the correct symmetry to break the frustrating $m_x$ mirror plane, we now study the optical phonons with a single crystal sample using Raman spectroscopy.  CoTi$_2$O$_5$ has 16 atoms in the primitive unit cell, giving 48 positional degrees of freedom. The $\Gamma$-point, 48-dimensional atomic displacement representation, $\Gamma_\mathrm{r}$, decomposes into the following irreducible representations;
$[\Gamma_\mathrm{r}=8\Gamma_{1}^{+} + 5\Gamma_{2}^{+} + 8\Gamma_{3}^{+} + 3\Gamma_{4}^{+} + 3\Gamma_{1}^{-} + 8\Gamma_{2}^{-} + 5\Gamma_{3}^{-} + 8\Gamma_{4}^{-} ]$. Of the 45 optic modes, 24 are Raman active (those that are parity even, as denoted by a $+$ subscript). Polarized Raman spectroscopy was performed in reflection geometry, with incident and scattered wavevectors parallel to the $c$-axis ($Z$). In this geometry only the 8 $\Gamma_1^+$ and 5 $\Gamma_2^+$ modes are excited. Furthermore, the $\Gamma_1^+$ excitations are isolated in the unrotated polarization channels $-Z(XX)Z$ and $-Z(YY)Z$ (in Porto's notation, where $X||a$ and $Y||b$), and the $\Gamma_2^+$ excitations are isolated in the rotated $-Z(XY)Z$ and $-Z(YX)Z$ channels. 

Figure \ref{fig:raman} $(a)$ shows the Raman spectra displaying all 5 $\Gamma_2^+$ modes with the same symmetry as the anomalous acoustic mode (related to $C_{66}$) discussed above.  The temperature dependence of select modes is presented in Figs. \ref{fig:raman} $(b-d)$.  Arguably, all $\Gamma_{2}^+$ modes display an anomaly at $T_{\rm N}$, however we note these effects are less than 0.025 meV which is negligible in comparison to the energetics of the $\Gamma_2^+$ acoustic phonon and magnons discussed above.  Fig. \ref{fig:raman}  $(c)$ illustrates a softening with temperature of the phonon mode near 21.5 meV that contrasts with the hardening typically expected with decreasing temperature and observed for the other optical phonons (see SI for temperature dependence of all 5 $\Gamma_{2}^{+}$ modes). Phonons typically harden in energy owing to loss of anharmonic effects and lattice contraction (Fig. \ref{fig:structural} $e$), and energetic softening supports a structural instability with the same symmetry analogous to soft optic phonon driven transitions in classic perovskites~\cite{Shirane74:46,Shirane70:2}.  This supports a structural instability at $T_{\rm N}$ with $\Gamma_{2}^{+}$ symmetry required to break the magnetic frustration outlined in Fig. \ref{fig:structural} $(b)$.


Density functional theory~\cite{Dudarev98:57,Liechtenstein95:52,Maksimov22:106,Streltsov05:71} was used to calculate the phonon eigenvectors and energies which were in agreement with the observed phonon mode energies (see SI). We highlight in the inset of Fig. \ref{fig:raman} $(a)$ the atomic displacements (eigenmode) associated with the $\sim$ 21.5 meV mode displayed in Fig. \ref{fig:raman} $(c)$. The displacements tied with this particular optic phonon mode move the oxygen atoms out of the mirror plane $m_{x}$.  This breaks the frustration illustrated in Fig. \ref{fig:structural} $(a,b)$.   We note that the low-energy acoustic shear mode, presented above (Fig. \ref{fig:phonons}), couples directly to the optic mode as it has the same $\Gamma_{2}^+$ symmetry~\cite{Cowley76:13} and importantly has the same energy as the dispersionless magnetic excitations along $K$.  This is suggestive of magnetoelastic coupling confirmed through recent macroscopic strain measurements that are able to switch magnetic domains in this compound.~\cite{Behr24:110}    

We observe an anomalous temperature dependence at $T_{\rm N}$ of an acoustic phonon mode with the correct symmetry to break the frustrating magnetic interactions illustrated in Fig. \ref{fig:structural} $(a,b)$ in the absence of an observable structural distortion with diffraction.  The energy scale of this anomaly (Fig. \ref{fig:phonons} $c$) is comparable to the energy scale of the magnetic exchange $J\sim$ 3 meV and $T_{\rm N}$, while significantly larger than the crystalline anisotropy $D\sim$ 0.25 meV discussed above.  This indicates a coupling between strain and magnetic order.  Supporting this and as noted in Ref. \onlinecite{Behr24:110}, considering structural and magnetic order parameters (defined by $\delta$ and $\Phi$, respectively), the lowest order invariant in the Landau expansion scales as $\sim \delta \Phi^{2}$ illustrating a phenomenological mechanism for coupling strain and magnetism in CoTi$_{2}$O$_{5}$.  Based on experimental and symmetry grounds, we propose that the magnetic transition in CoTi$_{2}$O$_{5}$ is dynamically driven through cooperative magnetoelastic fluctuations rather than conventional soft phonon dynamics (like in perovskites) or more exotic mechanisms such as nematic order or biquadratic exchange suggested in other triangular magnets~\cite{Stoudenmire09:79,Valentine20:6,Tsunetsugu06:75}. 

In summary, we report neutron inelastic scattering results that reveal a $\Gamma_2^+$ shear phonon that hardens at $T_{\rm N}$.  Furthermore, this phonon mode has the same symmetry as an anomalous optic phonon probed with Raman spectroscopy.   The $\Gamma_2^+$ acoustic shear mode energetics (in Fig. \ref{fig:phonons}) coincide with those of the magnons presented in Fig. \ref{fig:magnetic_dynamics}.  The absence of a measurable static structural distortion nor a magnetic dispersion along $K$ appears to discount a conventional magnetostructural displacive phase transition that breaks $m_x$.  Instead, we propose that our characterization of the dynamics implies that spin Jahn-Teller physics is driven by magnetoelastic dynamics.  We note that our results indicate the role of acoustic phonons in establishing the required magnetostructural coupling is at the heart of spin Jahn-Teller physics, when the acoustic phonon and magnon energetics are comparable and of the same symmetry.  Our results also suggest that acoustics are a promising avenue to investigate for the study of critical dynamics of magnetostructural transitions in materials where conventional soft lattice dynamics and changes in Bragg peaks are not observable.\\

The work done at UoE was funded by EPSRC (EP/V03605X/1, EP/R032130/1, EP/R013004/1) and the STFC. D.P. acknowledges EPSRC Grant No. EP/T028637/1, and the Oxford-ShanhagaiTech Collaboration project for financial support. Work at Cardiff was funded through EP/X034739/1. S.J.B. acknowledges support from UK Research and Innovation (UKRI) under the UK government’s Horizon Europe funding guarantee (Grant No. EP/X025861/1). D.F.T. calculations were supported by the Russian Science Foundation (Project No. RSF 23-12-00159), while their analysis by Ministry of Science and Higher Education through project Quantum.


\begin{thebibliography}{57}%
	\makeatletter
	\providecommand \@ifxundefined [1]{%
		\@ifx{#1\undefined}
	}%
	\providecommand \@ifnum [1]{%
		\ifnum #1\expandafter \@firstoftwo
		\else \expandafter \@secondoftwo
		\fi
	}%
	\providecommand \@ifx [1]{%
		\ifx #1\expandafter \@firstoftwo
		\else \expandafter \@secondoftwo
		\fi
	}%
	\providecommand \natexlab [1]{#1}%
	\providecommand \enquote  [1]{``#1''}%
	\providecommand \bibnamefont  [1]{#1}%
	\providecommand \bibfnamefont [1]{#1}%
	\providecommand \citenamefont [1]{#1}%
	\providecommand \href@noop [0]{\@secondoftwo}%
	\providecommand \href [0]{\begingroup \@sanitize@url \@href}%
	\providecommand \@href[1]{\@@startlink{#1}\@@href}%
	\providecommand \@@href[1]{\endgroup#1\@@endlink}%
	\providecommand \@sanitize@url [0]{\catcode `\\12\catcode `\$12\catcode
		`\&12\catcode `\#12\catcode `\^12\catcode `\_12\catcode `\%12\relax}%
	\providecommand \@@startlink[1]{}%
	\providecommand \@@endlink[0]{}%
	\providecommand \url  [0]{\begingroup\@sanitize@url \@url }%
	\providecommand \@url [1]{\endgroup\@href {#1}{\urlprefix }}%
	\providecommand \urlprefix  [0]{URL }%
	\providecommand \Eprint [0]{\href }%
	\providecommand \doibase [0]{https://doi.org/}%
	\providecommand \selectlanguage [0]{\@gobble}%
	\providecommand \bibinfo  [0]{\@secondoftwo}%
	\providecommand \bibfield  [0]{\@secondoftwo}%
	\providecommand \translation [1]{[#1]}%
	\providecommand \BibitemOpen [0]{}%
	\providecommand \bibitemStop [0]{}%
	\providecommand \bibitemNoStop [0]{.\EOS\space}%
	\providecommand \EOS [0]{\spacefactor3000\relax}%
	\providecommand \BibitemShut  [1]{\csname bibitem#1\endcsname}%
	\let\auto@bib@innerbib\@empty
	\bibitem [{\citenamefont {Cochran}(1960)}]{Cochran60:9}%
	\BibitemOpen
	\bibfield  {author} {\bibinfo {author} {\bibfnamefont {W.}~\bibnamefont
			{Cochran}},\ }\bibfield  {title} {\bibinfo {title} {Crystal stability and the
			theory of ferroelectricity},\ }\href
	{https://doi.org/10.1080/00018736000101229} {\bibfield  {journal} {\bibinfo
			{journal} {Adv. Phys.}\ }\textbf {\bibinfo {volume} {9}},\ \bibinfo {pages}
		{387} (\bibinfo {year} {1960})}\BibitemShut {NoStop}%
	\bibitem [{\citenamefont {Cochran}(1961)}]{Cochran61:10}%
	\BibitemOpen
	\bibfield  {author} {\bibinfo {author} {\bibfnamefont {W.}~\bibnamefont
			{Cochran}},\ }\bibfield  {title} {\bibinfo {title} {Crystal stability and the
			theory of ferroelectricity .2. piezoelectric crystals},\ }\href
	{https://doi.org/10.1080/00018736100101321} {\bibfield  {journal} {\bibinfo
			{journal} {Adv. Phys.}\ }\textbf {\bibinfo {volume} {10}},\ \bibinfo {pages}
		{401} (\bibinfo {year} {1961})}\BibitemShut {NoStop}%
	\bibitem [{\citenamefont {Shirane}(1974)}]{Shirane74:46}%
	\BibitemOpen
	\bibfield  {author} {\bibinfo {author} {\bibfnamefont {G.}~\bibnamefont
			{Shirane}},\ }\bibfield  {title} {\bibinfo {title} {Neutron scattering
			studies of structural phase transitions at \textsc{B}rookhaven},\ }\href
	{https://doi.org/10.1103/RevModPhys.46.437} {\bibfield  {journal} {\bibinfo
			{journal} {Rev. Mod. Phys.}\ }\textbf {\bibinfo {volume} {46}},\ \bibinfo
		{pages} {437} (\bibinfo {year} {1974})}\BibitemShut {NoStop}%
	\bibitem [{\citenamefont {Cowley}\ and\ \citenamefont
		{Shapiro}(2006)}]{Cowley06:75}%
	\BibitemOpen
	\bibfield  {author} {\bibinfo {author} {\bibfnamefont {R.~A.}\ \bibnamefont
			{Cowley}}\ and\ \bibinfo {author} {\bibfnamefont {S.~M.}\ \bibnamefont
			{Shapiro}},\ }\bibfield  {title} {\bibinfo {title} {Structural phase
			transitions},\ }\href {https://doi.org/10.1143/JPSJ.75.111001} {\bibfield
		{journal} {\bibinfo  {journal} {J. Phys. Soc. Jpn.}\ }\textbf {\bibinfo
			{volume} {75}},\ \bibinfo {pages} {111001} (\bibinfo {year}
		{2006})}\BibitemShut {NoStop}%
	\bibitem [{\citenamefont {Shirane}\ \emph {et~al.}(1970)\citenamefont
		{Shirane}, \citenamefont {Axe}, \citenamefont {Harada},\ and\ \citenamefont
		{Remeika}}]{Shirane70:2}%
	\BibitemOpen
	\bibfield  {author} {\bibinfo {author} {\bibfnamefont {G.}~\bibnamefont
			{Shirane}}, \bibinfo {author} {\bibfnamefont {J.~D.}\ \bibnamefont {Axe}},
		\bibinfo {author} {\bibfnamefont {J.}~\bibnamefont {Harada}},\ and\ \bibinfo
		{author} {\bibfnamefont {J.~P.}\ \bibnamefont {Remeika}},\ }\bibfield
	{title} {\bibinfo {title} {Soft ferroelectric modes in lead titanate},\
	}\href {https://doi.org/10.1103/PhysRevB.2.155} {\bibfield  {journal}
		{\bibinfo  {journal} {Phys. Rev. B}\ }\textbf {\bibinfo {volume} {2}},\
		\bibinfo {pages} {155} (\bibinfo {year} {1970})}\BibitemShut {NoStop}%
	\bibitem [{\citenamefont {Harada}\ \emph {et~al.}(1970)\citenamefont {Harada},
		\citenamefont {Axe},\ and\ \citenamefont {Shirane}}]{Harada70:608}%
	\BibitemOpen
	\bibfield  {author} {\bibinfo {author} {\bibfnamefont {J.}~\bibnamefont
			{Harada}}, \bibinfo {author} {\bibfnamefont {J.~D.}\ \bibnamefont {Axe}},\
		and\ \bibinfo {author} {\bibfnamefont {G.}~\bibnamefont {Shirane}},\
	}\bibfield  {title} {\bibinfo {title} {Determination of the normal
			vibrational displacements in several perovskites by inelastic neutron
			scattering},\ }\href {https://doi.org/10.1107/S0567739470001547} {\bibfield
		{journal} {\bibinfo  {journal} {Acta Cryst.}\ }\textbf {\bibinfo {volume}
			{A26}},\ \bibinfo {pages} {608} (\bibinfo {year} {1970})}\BibitemShut
	{NoStop}%
	\bibitem [{\citenamefont {Harada}\ \emph {et~al.}(1971)\citenamefont {Harada},
		\citenamefont {Axe},\ and\ \citenamefont {Shirane}}]{Harada71:4}%
	\BibitemOpen
	\bibfield  {author} {\bibinfo {author} {\bibfnamefont {J.}~\bibnamefont
			{Harada}}, \bibinfo {author} {\bibfnamefont {J.~D.}\ \bibnamefont {Axe}},\
		and\ \bibinfo {author} {\bibfnamefont {G.}~\bibnamefont {Shirane}},\
	}\bibfield  {title} {\bibinfo {title} {Neutron-scattering study of soft modes
			in cubic \textsc{B}a\textsc{T}i\textsc{O}$_3$},\ }\href
	{https://doi.org/10.1103/PhysRevB.4.155} {\bibfield  {journal} {\bibinfo
			{journal} {Phys. Rev. B}\ }\textbf {\bibinfo {volume} {4}},\ \bibinfo {pages}
		{155} (\bibinfo {year} {1971})}\BibitemShut {NoStop}%
	\bibitem [{\citenamefont {Yamada}\ \emph {et~al.}(1974)\citenamefont {Yamada},
		\citenamefont {Noda}, \citenamefont {Axe},\ and\ \citenamefont
		{Shirane}}]{Yamada74:9}%
	\BibitemOpen
	\bibfield  {author} {\bibinfo {author} {\bibfnamefont {Y.}~\bibnamefont
			{Yamada}}, \bibinfo {author} {\bibfnamefont {Y.}~\bibnamefont {Noda}},
		\bibinfo {author} {\bibfnamefont {J.~D.}\ \bibnamefont {Axe}},\ and\ \bibinfo
		{author} {\bibfnamefont {G.}~\bibnamefont {Shirane}},\ }\bibfield  {title}
	{\bibinfo {title} {Dynamical critical phenomena in
			\textsc{N}\textsc{D}$_{4}$\textsc{B}r},\ }\href
	{https://doi.org/10.1103/PhysRevB.9.4429} {\bibfield  {journal} {\bibinfo
			{journal} {Phys. Rev. B}\ }\textbf {\bibinfo {volume} {9}},\ \bibinfo {pages}
		{4429} (\bibinfo {year} {1974})}\BibitemShut {NoStop}%
	\bibitem [{\citenamefont {Harris}\ \emph {et~al.}(1998)\citenamefont {Harris},
		\citenamefont {Dove}, \citenamefont {Swainson},\ and\ \citenamefont
		{Hagen}}]{Harris98:10}%
	\BibitemOpen
	\bibfield  {author} {\bibinfo {author} {\bibfnamefont {M.~J.}\ \bibnamefont
			{Harris}}, \bibinfo {author} {\bibfnamefont {M.~T.}\ \bibnamefont {Dove}},
		\bibinfo {author} {\bibfnamefont {I.~P.}\ \bibnamefont {Swainson}},\ and\
		\bibinfo {author} {\bibfnamefont {M.~E.}\ \bibnamefont {Hagen}},\ }\bibfield
	{title} {\bibinfo {title} {Anomalous dynamical effects in calcite
			\textsc{C}a\textsc{C}\textsc{O}$_3$},\ }\href
	{https://doi.org/10.1088/0953-8984/10/25/002} {\bibfield  {journal} {\bibinfo
			{journal} {J. Phys.: Condens. Matter}\ }\textbf {\bibinfo {volume} {10}},\
		\bibinfo {pages} {L423} (\bibinfo {year} {1998})}\BibitemShut {NoStop}%
	\bibitem [{\citenamefont {Kenzelmann}\ \emph {et~al.}(2007)\citenamefont
		{Kenzelmann}, \citenamefont {Lawes}, \citenamefont {Harris}, \citenamefont
		{Gasparovic}, \citenamefont {Broholm}, \citenamefont {Ramirez}, \citenamefont
		{Jorge}, \citenamefont {Jaime}, \citenamefont {Park}, \citenamefont {Huang},
		\citenamefont {Shapiro},\ and\ \citenamefont {Demianets}}]{Kenzelmann07:98}%
	\BibitemOpen
	\bibfield  {author} {\bibinfo {author} {\bibfnamefont {M.}~\bibnamefont
			{Kenzelmann}}, \bibinfo {author} {\bibfnamefont {G.}~\bibnamefont {Lawes}},
		\bibinfo {author} {\bibfnamefont {A.~B.}\ \bibnamefont {Harris}}, \bibinfo
		{author} {\bibfnamefont {G.}~\bibnamefont {Gasparovic}}, \bibinfo {author}
		{\bibfnamefont {C.}~\bibnamefont {Broholm}}, \bibinfo {author} {\bibfnamefont
			{A.~P.}\ \bibnamefont {Ramirez}}, \bibinfo {author} {\bibfnamefont {G.~A.}\
			\bibnamefont {Jorge}}, \bibinfo {author} {\bibfnamefont {M.}~\bibnamefont
			{Jaime}}, \bibinfo {author} {\bibfnamefont {S.}~\bibnamefont {Park}},
		\bibinfo {author} {\bibfnamefont {Q.}~\bibnamefont {Huang}}, \bibinfo
		{author} {\bibfnamefont {A.~Y.}\ \bibnamefont {Shapiro}},\ and\ \bibinfo
		{author} {\bibfnamefont {L.~A.}\ \bibnamefont {Demianets}},\ }\bibfield
	{title} {\bibinfo {title} {Direct transition from a disordered to a
			multiferroic phase on a triangular lattice},\ }\href
	{https://doi.org/10.1103/PhysRevLett.98.267205} {\bibfield  {journal}
		{\bibinfo  {journal} {Phys. Rev. Lett.}\ }\textbf {\bibinfo {volume} {98}},\
		\bibinfo {pages} {267205} (\bibinfo {year} {2007})}\BibitemShut {NoStop}%
	\bibitem [{\citenamefont {Hearmon}\ \emph {et~al.}(2012)\citenamefont
		{Hearmon}, \citenamefont {Fabrizi}, \citenamefont {Chapon}, \citenamefont
		{Johnson}, \citenamefont {Prabhakaran}, \citenamefont {Streltsov},
		\citenamefont {Brown},\ and\ \citenamefont {Radaelli}}]{Hearmon12:108}%
	\BibitemOpen
	\bibfield  {author} {\bibinfo {author} {\bibfnamefont {A.~J.}\ \bibnamefont
			{Hearmon}}, \bibinfo {author} {\bibfnamefont {F.}~\bibnamefont {Fabrizi}},
		\bibinfo {author} {\bibfnamefont {L.~C.}\ \bibnamefont {Chapon}}, \bibinfo
		{author} {\bibfnamefont {R.~D.}\ \bibnamefont {Johnson}}, \bibinfo {author}
		{\bibfnamefont {D.}~\bibnamefont {Prabhakaran}}, \bibinfo {author}
		{\bibfnamefont {S.~V.}\ \bibnamefont {Streltsov}}, \bibinfo {author}
		{\bibfnamefont {P.~J.}\ \bibnamefont {Brown}},\ and\ \bibinfo {author}
		{\bibfnamefont {P.~G.}\ \bibnamefont {Radaelli}},\ }\bibfield  {title}
	{\bibinfo {title} {Electric field control of the magnetic chiralities in
			ferroaxial multiferroic
			\textsc{R}b\textsc{F}e(\textsc{M}o\textsc{O}$_4$)$_2$},\ }\href
	{https://doi.org/10.1103/PhysRevLett.108.237201} {\bibfield  {journal}
		{\bibinfo  {journal} {Phys. Rev. Lett.}\ }\textbf {\bibinfo {volume} {108}},\
		\bibinfo {pages} {237201} (\bibinfo {year} {2012})}\BibitemShut {NoStop}%
	\bibitem [{\citenamefont {Johnson}\ \emph {et~al.}(2012)\citenamefont
		{Johnson}, \citenamefont {Chapon}, \citenamefont {Khalyavin}, \citenamefont
		{Manuel}, \citenamefont {Radaelli},\ and\ \citenamefont
		{Martin}}]{Johnson12:108}%
	\BibitemOpen
	\bibfield  {author} {\bibinfo {author} {\bibfnamefont {R.~D.}\ \bibnamefont
			{Johnson}}, \bibinfo {author} {\bibfnamefont {L.~C.}\ \bibnamefont {Chapon}},
		\bibinfo {author} {\bibfnamefont {D.~D.}\ \bibnamefont {Khalyavin}}, \bibinfo
		{author} {\bibfnamefont {P.}~\bibnamefont {Manuel}}, \bibinfo {author}
		{\bibfnamefont {P.~G.}\ \bibnamefont {Radaelli}},\ and\ \bibinfo {author}
		{\bibfnamefont {C.}~\bibnamefont {Martin}},\ }\bibfield  {title} {\bibinfo
		{title} {Giant improper ferroelectricity in the ferroaxial magnet
			\textsc{C}a\textsc{M}n$_{7}$\textsc{O}$_{12}$},\ }\href
	{https://doi.org/10.1103/PhysRevLett.108.067201} {\bibfield  {journal}
		{\bibinfo  {journal} {Phys. Rev. Lett.}\ }\textbf {\bibinfo {volume} {108}},\
		\bibinfo {pages} {067201} (\bibinfo {year} {2012})}\BibitemShut {NoStop}%
	\bibitem [{\citenamefont {Kadlec}\ \emph {et~al.}(2014)\citenamefont {Kadlec},
		\citenamefont {Goian}, \citenamefont {Kadlec}, \citenamefont {Kempa},
		\citenamefont {Van\ifmmode~\check{e}\else \v{e}\fi{}k}, \citenamefont
		{Taylor}, \citenamefont {Rols}, \citenamefont {Prokle\ifmmode~\check{s}\else
			\v{s}\fi{}ka}, \citenamefont {Orlita},\ and\ \citenamefont
		{Kamba}}]{Kadlec14:90}%
	\BibitemOpen
	\bibfield  {author} {\bibinfo {author} {\bibfnamefont {F.}~\bibnamefont
			{Kadlec}}, \bibinfo {author} {\bibfnamefont {V.}~\bibnamefont {Goian}},
		\bibinfo {author} {\bibfnamefont {C.}~\bibnamefont {Kadlec}}, \bibinfo
		{author} {\bibfnamefont {M.}~\bibnamefont {Kempa}}, \bibinfo {author}
		{\bibfnamefont {P.~c.~v.}\ \bibnamefont {Van\ifmmode~\check{e}\else
				\v{e}\fi{}k}}, \bibinfo {author} {\bibfnamefont {J.}~\bibnamefont {Taylor}},
		\bibinfo {author} {\bibfnamefont {S.}~\bibnamefont {Rols}}, \bibinfo {author}
		{\bibfnamefont {J.}~\bibnamefont {Prokle\ifmmode~\check{s}\else
				\v{s}\fi{}ka}}, \bibinfo {author} {\bibfnamefont {M.}~\bibnamefont
			{Orlita}},\ and\ \bibinfo {author} {\bibfnamefont {S.}~\bibnamefont
			{Kamba}},\ }\bibfield  {title} {\bibinfo {title} {Possible coupling between
			magnons and phonons in multiferroic
			\textsc{C}a\textsc{M}n$_7$\textsc{O}$_{12}$},\ }\href
	{https://doi.org/10.1103/PhysRevB.90.054307} {\bibfield  {journal} {\bibinfo
			{journal} {Phys. Rev. B}\ }\textbf {\bibinfo {volume} {90}},\ \bibinfo
		{pages} {054307} (\bibinfo {year} {2014})}\BibitemShut {NoStop}%
	\bibitem [{\citenamefont {Johnson}\ \emph {et~al.}(2011)\citenamefont
		{Johnson}, \citenamefont {Nair}, \citenamefont {Chapon}, \citenamefont
		{Bombardi}, \citenamefont {Vecchini}, \citenamefont {Prabhakaran},
		\citenamefont {Boothroyd},\ and\ \citenamefont {Radaelli}}]{Johnson11:107}%
	\BibitemOpen
	\bibfield  {author} {\bibinfo {author} {\bibfnamefont {R.~D.}\ \bibnamefont
			{Johnson}}, \bibinfo {author} {\bibfnamefont {S.}~\bibnamefont {Nair}},
		\bibinfo {author} {\bibfnamefont {L.~C.}\ \bibnamefont {Chapon}}, \bibinfo
		{author} {\bibfnamefont {A.}~\bibnamefont {Bombardi}}, \bibinfo {author}
		{\bibfnamefont {C.}~\bibnamefont {Vecchini}}, \bibinfo {author}
		{\bibfnamefont {D.}~\bibnamefont {Prabhakaran}}, \bibinfo {author}
		{\bibfnamefont {A.~T.}\ \bibnamefont {Boothroyd}},\ and\ \bibinfo {author}
		{\bibfnamefont {P.~G.}\ \bibnamefont {Radaelli}},\ }\bibfield  {title}
	{\bibinfo {title} {\textsc{C}u$_{3}$\textsc{N}b$_{2}$\textsc{O}$_{8}$: A
			multiferroic with chiral coupling to the crystal structure},\ }\href
	{https://doi.org/10.1103/PhysRevLett.107.137205} {\bibfield  {journal}
		{\bibinfo  {journal} {Phys. Rev. Lett.}\ }\textbf {\bibinfo {volume} {107}},\
		\bibinfo {pages} {137205} (\bibinfo {year} {2011})}\BibitemShut {NoStop}%
	\bibitem [{\citenamefont {Giles-Donovan}\ \emph {et~al.}(2020)\citenamefont
		{Giles-Donovan}, \citenamefont {Qureshi}, \citenamefont {Johnson},
		\citenamefont {Zhang}, \citenamefont {Cheong}, \citenamefont {Cochran},\ and\
		\citenamefont {Stock}}]{Giles-Donovan20:102}%
	\BibitemOpen
	\bibfield  {author} {\bibinfo {author} {\bibfnamefont {N.}~\bibnamefont
			{Giles-Donovan}}, \bibinfo {author} {\bibfnamefont {N.}~\bibnamefont
			{Qureshi}}, \bibinfo {author} {\bibfnamefont {R.~D.}\ \bibnamefont
			{Johnson}}, \bibinfo {author} {\bibfnamefont {L.~Y.}\ \bibnamefont {Zhang}},
		\bibinfo {author} {\bibfnamefont {S.-W.}\ \bibnamefont {Cheong}}, \bibinfo
		{author} {\bibfnamefont {S.}~\bibnamefont {Cochran}},\ and\ \bibinfo {author}
		{\bibfnamefont {C.}~\bibnamefont {Stock}},\ }\bibfield  {title} {\bibinfo
		{title} {Imitation of spin density wave order in
			\textsc{C}u$_3$\textsc{N}b$_2$\textsc{O}$_8$},\ }\href
	{https://doi.org/10.1103/PhysRevB.102.024414} {\bibfield  {journal} {\bibinfo
			{journal} {Phys. Rev. B}\ }\textbf {\bibinfo {volume} {102}},\ \bibinfo
		{pages} {024414} (\bibinfo {year} {2020})}\BibitemShut {NoStop}%
	\bibitem [{\citenamefont {Kamba}(2021)}]{Kambda21:9}%
	\BibitemOpen
	\bibfield  {author} {\bibinfo {author} {\bibfnamefont {S.}~\bibnamefont
			{Kamba}},\ }\bibfield  {title} {\bibinfo {title} {Soft-mode spectroscopy of
			ferroelectrics and multiferroics: \textsc{A} review},\ }\href
	{https://doi.org/doi.org/10.1063/5.0036066} {\bibfield  {journal} {\bibinfo
			{journal} {APL Mater.}\ }\textbf {\bibinfo {volume} {9}},\ \bibinfo {pages}
		{020704} (\bibinfo {year} {2021})}\BibitemShut {NoStop}%
	\bibitem [{\citenamefont {Ding}\ \emph {et~al.}(2023)\citenamefont {Ding},
		\citenamefont {Colin}, \citenamefont {Simonet}, \citenamefont {Stock},
		\citenamefont {Brubach}, \citenamefont {Verseils}, \citenamefont {Roy},
		\citenamefont {Sakai}, \citenamefont {Koza}, \citenamefont {Piovano},
		\citenamefont {Ivanov}, \citenamefont {Rodriguez-Rivera}, \citenamefont
		{de~Brion},\ and\ \citenamefont {Songvilay}}]{Ding23:7}%
	\BibitemOpen
	\bibfield  {author} {\bibinfo {author} {\bibfnamefont {L.}~\bibnamefont
			{Ding}}, \bibinfo {author} {\bibfnamefont {C.~V.}\ \bibnamefont {Colin}},
		\bibinfo {author} {\bibfnamefont {V.}~\bibnamefont {Simonet}}, \bibinfo
		{author} {\bibfnamefont {C.}~\bibnamefont {Stock}}, \bibinfo {author}
		{\bibfnamefont {J.-B.}\ \bibnamefont {Brubach}}, \bibinfo {author}
		{\bibfnamefont {M.}~\bibnamefont {Verseils}}, \bibinfo {author}
		{\bibfnamefont {P.}~\bibnamefont {Roy}}, \bibinfo {author} {\bibfnamefont
			{V.~G.}\ \bibnamefont {Sakai}}, \bibinfo {author} {\bibfnamefont {M.~M.}\
			\bibnamefont {Koza}}, \bibinfo {author} {\bibfnamefont {A.}~\bibnamefont
			{Piovano}}, \bibinfo {author} {\bibfnamefont {A.}~\bibnamefont {Ivanov}},
		\bibinfo {author} {\bibfnamefont {J.~A.}\ \bibnamefont {Rodriguez-Rivera}},
		\bibinfo {author} {\bibfnamefont {S.}~\bibnamefont {de~Brion}},\ and\
		\bibinfo {author} {\bibfnamefont {M.}~\bibnamefont {Songvilay}},\ }\bibfield
	{title} {\bibinfo {title} {Lattice dynamics and spin excitations in the
			metal-organic framework
			\textsc{CH}$_{3}$\textsc{NH}$_{3}$\textsc{C}o\textsc{HCOO}$_3$},\ }\href
	{https://doi.org/10.1103/PhysRevMaterials.7.084405} {\bibfield  {journal}
		{\bibinfo  {journal} {Phys. Rev. Mater.}\ }\textbf {\bibinfo {volume} {7}},\
		\bibinfo {pages} {084405} (\bibinfo {year} {2023})}\BibitemShut {NoStop}%
	\bibitem [{\citenamefont {Gehring}\ and\ \citenamefont
		{Gehring}(1975)}]{Gehring75:38}%
	\BibitemOpen
	\bibfield  {author} {\bibinfo {author} {\bibfnamefont {G.~A.}\ \bibnamefont
			{Gehring}}\ and\ \bibinfo {author} {\bibfnamefont {K.~A.}\ \bibnamefont
			{Gehring}},\ }\bibfield  {title} {\bibinfo {title} {Co-operative
			\textsc{J}ahn-\textsc{T}eller effects},\ }\href
	{https://doi.org/10.1088/0034-4885/38/1/001} {\bibfield  {journal} {\bibinfo
			{journal} {Rep. Prog. Phys.}\ }\textbf {\bibinfo {volume} {38}},\ \bibinfo
		{pages} {1} (\bibinfo {year} {1975})}\BibitemShut {NoStop}%
	\bibitem [{\citenamefont {Tchernyshyov}\ \emph {et~al.}(2002)\citenamefont
		{Tchernyshyov}, \citenamefont {Moessner},\ and\ \citenamefont
		{Sondhi}}]{Tchernyshyov02:88}%
	\BibitemOpen
	\bibfield  {author} {\bibinfo {author} {\bibfnamefont {O.}~\bibnamefont
			{Tchernyshyov}}, \bibinfo {author} {\bibfnamefont {R.}~\bibnamefont
			{Moessner}},\ and\ \bibinfo {author} {\bibfnamefont {S.~L.}\ \bibnamefont
			{Sondhi}},\ }\bibfield  {title} {\bibinfo {title} {Order by distortion and
			string modes in pyrochlore antiferromagnets},\ }\href
	{https://doi.org/10.1103/PhysRevLett.88.067203} {\bibfield  {journal}
		{\bibinfo  {journal} {Phys. Rev. Lett.}\ }\textbf {\bibinfo {volume} {88}},\
		\bibinfo {pages} {067203} (\bibinfo {year} {2002})}\BibitemShut {NoStop}%
	\bibitem [{\citenamefont {Yamashita}\ and\ \citenamefont
		{Ueda}(2000)}]{Yamashita00:85}%
	\BibitemOpen
	\bibfield  {author} {\bibinfo {author} {\bibfnamefont {Y.}~\bibnamefont
			{Yamashita}}\ and\ \bibinfo {author} {\bibfnamefont {K.}~\bibnamefont
			{Ueda}},\ }\bibfield  {title} {\bibinfo {title} {Spin-driven
			\textsc{J}ahn-\textsc{T}eller distortion in a pyrochlore system},\ }\href
	{https://doi.org/10.1103/PhysRevLett.85.4960} {\bibfield  {journal} {\bibinfo
			{journal} {Phys. Rev. Lett.}\ }\textbf {\bibinfo {volume} {85}},\ \bibinfo
		{pages} {4960} (\bibinfo {year} {2000})}\BibitemShut {NoStop}%
	\bibitem [{\citenamefont {Mullerbuschbaum}\ and\ \citenamefont
		{Waburg}(1983)}]{Muller83:114}%
	\BibitemOpen
	\bibfield  {author} {\bibinfo {author} {\bibfnamefont {H.}~\bibnamefont
			{Mullerbuschbaum}}\ and\ \bibinfo {author} {\bibfnamefont {M.}~\bibnamefont
			{Waburg}},\ }\bibfield  {title} {\bibinfo {title} {Pseudobrookite mit
			weitgehend geordneter metallverteilung:
			\textsc{C}o\textsc{T}i$_{2}$\textsc{0}$_{5}$,
			\textsc{M}g\textsc{T}i$_{2}$\textsc{0}$_{5}$ und
			\textsc{F}e\textsc{T}i$_{2}$\textsc{0}$_{5}$},\ }\href
	{https://doi.org/10.1007/BF00809371} {\bibfield  {journal} {\bibinfo
			{journal} {Monatsch. Chem.}\ }\textbf {\bibinfo {volume} {114}},\ \bibinfo
		{pages} {21} (\bibinfo {year} {1983})}\BibitemShut {NoStop}%
	\bibitem [{\citenamefont {Streltsov}\ and\ \citenamefont
		{Khomskii}(2020)}]{Streltsov20:10}%
	\BibitemOpen
	\bibfield  {author} {\bibinfo {author} {\bibfnamefont {S.~V.}\ \bibnamefont
			{Streltsov}}\ and\ \bibinfo {author} {\bibfnamefont {D.~I.}\ \bibnamefont
			{Khomskii}},\ }\bibfield  {title} {\bibinfo {title}
		{\textsc{J}ahn-\textsc{T}eller effect and spin-orbit coupling: Friends or
			foes?},\ }\href {https://doi.org/10.1103/PhysRevX.10.031043} {\bibfield
		{journal} {\bibinfo  {journal} {Phys. Rev. X}\ }\textbf {\bibinfo {volume}
			{10}},\ \bibinfo {pages} {031043} (\bibinfo {year} {2020})}\BibitemShut
	{NoStop}%
	\bibitem [{\citenamefont {Cowley}\ \emph {et~al.}(2013)\citenamefont {Cowley},
		\citenamefont {Buyers}, \citenamefont {Stock}, \citenamefont {Yamani},
		\citenamefont {Frost}, \citenamefont {Taylor},\ and\ \citenamefont
		{Prabhakaran}}]{Cowley13:88}%
	\BibitemOpen
	\bibfield  {author} {\bibinfo {author} {\bibfnamefont {R.~A.}\ \bibnamefont
			{Cowley}}, \bibinfo {author} {\bibfnamefont {W.~J.~L.}\ \bibnamefont
			{Buyers}}, \bibinfo {author} {\bibfnamefont {C.}~\bibnamefont {Stock}},
		\bibinfo {author} {\bibfnamefont {Z.}~\bibnamefont {Yamani}}, \bibinfo
		{author} {\bibfnamefont {C.}~\bibnamefont {Frost}}, \bibinfo {author}
		{\bibfnamefont {J.~W.}\ \bibnamefont {Taylor}},\ and\ \bibinfo {author}
		{\bibfnamefont {D.}~\bibnamefont {Prabhakaran}},\ }\bibfield  {title}
	{\bibinfo {title} {Neutron scattering investigation of the $d\ensuremath{-}d$
			excitations below the mott gap of \textsc{C}o\textsc{O}},\ }\href
	{https://doi.org/10.1103/PhysRevB.88.205117} {\bibfield  {journal} {\bibinfo
			{journal} {Phys. Rev. B}\ }\textbf {\bibinfo {volume} {88}},\ \bibinfo
		{pages} {205117} (\bibinfo {year} {2013})}\BibitemShut {NoStop}%
	\bibitem [{\citenamefont {Wallington}\ \emph {et~al.}(2015)\citenamefont
		{Wallington}, \citenamefont {Arevalo-Lopez}, \citenamefont {Taylor},
		\citenamefont {Stewart}, \citenamefont {Garcia-Sakai}, \citenamefont
		{Attfield},\ and\ \citenamefont {Stock}}]{Wallington15:92}%
	\BibitemOpen
	\bibfield  {author} {\bibinfo {author} {\bibfnamefont {F.}~\bibnamefont
			{Wallington}}, \bibinfo {author} {\bibfnamefont {A.~M.}\ \bibnamefont
			{Arevalo-Lopez}}, \bibinfo {author} {\bibfnamefont {J.~W.}\ \bibnamefont
			{Taylor}}, \bibinfo {author} {\bibfnamefont {J.~R.}\ \bibnamefont {Stewart}},
		\bibinfo {author} {\bibfnamefont {V.}~\bibnamefont {Garcia-Sakai}}, \bibinfo
		{author} {\bibfnamefont {J.~P.}\ \bibnamefont {Attfield}},\ and\ \bibinfo
		{author} {\bibfnamefont {C.}~\bibnamefont {Stock}},\ }\bibfield  {title}
	{\bibinfo {title} {Spin-orbit transitions in $\ensuremath{\alpha}$- and
			$\ensuremath{\gamma}$-\textsc{C}o\textsc{V}$_{2}$\textsc{O}$_{6}$},\ }\href
	{https://doi.org/10.1103/PhysRevB.92.125116} {\bibfield  {journal} {\bibinfo
			{journal} {Phys. Rev. B}\ }\textbf {\bibinfo {volume} {92}},\ \bibinfo
		{pages} {125116} (\bibinfo {year} {2015})}\BibitemShut {NoStop}%
	\bibitem [{\citenamefont {Sarte}\ \emph
		{et~al.}(2018{\natexlab{a}})\citenamefont {Sarte}, \citenamefont {Cowley},
		\citenamefont {Rodriguez}, \citenamefont {Pachoud}, \citenamefont {Le},
		\citenamefont {Garc\'{\i}a-Sakai}, \citenamefont {Taylor}, \citenamefont
		{Frost}, \citenamefont {Prabhakaran}, \citenamefont {MacEwen}, \citenamefont
		{Kitada}, \citenamefont {Browne}, \citenamefont {Songvilay}, \citenamefont
		{Yamani}, \citenamefont {Buyers}, \citenamefont {Attfield},\ and\
		\citenamefont {Stock}}]{Sarte18:98}%
	\BibitemOpen
	\bibfield  {author} {\bibinfo {author} {\bibfnamefont {P.~M.}\ \bibnamefont
			{Sarte}}, \bibinfo {author} {\bibfnamefont {R.~A.}\ \bibnamefont {Cowley}},
		\bibinfo {author} {\bibfnamefont {E.~E.}\ \bibnamefont {Rodriguez}}, \bibinfo
		{author} {\bibfnamefont {E.}~\bibnamefont {Pachoud}}, \bibinfo {author}
		{\bibfnamefont {D.}~\bibnamefont {Le}}, \bibinfo {author} {\bibfnamefont
			{V.}~\bibnamefont {Garc\'{\i}a-Sakai}}, \bibinfo {author} {\bibfnamefont
			{J.~W.}\ \bibnamefont {Taylor}}, \bibinfo {author} {\bibfnamefont {C.~D.}\
			\bibnamefont {Frost}}, \bibinfo {author} {\bibfnamefont {D.}~\bibnamefont
			{Prabhakaran}}, \bibinfo {author} {\bibfnamefont {C.}~\bibnamefont
			{MacEwen}}, \bibinfo {author} {\bibfnamefont {A.}~\bibnamefont {Kitada}},
		\bibinfo {author} {\bibfnamefont {A.~J.}\ \bibnamefont {Browne}}, \bibinfo
		{author} {\bibfnamefont {M.}~\bibnamefont {Songvilay}}, \bibinfo {author}
		{\bibfnamefont {Z.}~\bibnamefont {Yamani}}, \bibinfo {author} {\bibfnamefont
			{W.~J.~L.}\ \bibnamefont {Buyers}}, \bibinfo {author} {\bibfnamefont {J.~P.}\
			\bibnamefont {Attfield}},\ and\ \bibinfo {author} {\bibfnamefont
			{C.}~\bibnamefont {Stock}},\ }\bibfield  {title} {\bibinfo {title}
		{Disentangling orbital and spin exchange interactions for \textsc{C}o$^{2+}$
			on a rocksalt lattice},\ }\href {https://doi.org/10.1103/PhysRevB.98.024415}
	{\bibfield  {journal} {\bibinfo  {journal} {Phys. Rev. B}\ }\textbf {\bibinfo
			{volume} {98}},\ \bibinfo {pages} {024415} (\bibinfo {year}
		{2018}{\natexlab{a}})}\BibitemShut {NoStop}%
	\bibitem [{\citenamefont {Sarte}\ \emph
		{et~al.}(2018{\natexlab{b}})\citenamefont {Sarte}, \citenamefont
		{Ar\'evalo-L\'opez}, \citenamefont {Songvilay}, \citenamefont {Le},
		\citenamefont {Guidi}, \citenamefont {Garc\'{\i}a-Sakai}, \citenamefont
		{Mukhopadhyay}, \citenamefont {Capelli}, \citenamefont {Ratcliff},
		\citenamefont {Hong}, \citenamefont {McNally}, \citenamefont {Pachoud},
		\citenamefont {Attfield},\ and\ \citenamefont {Stock}}]{Sarte18:98_2}%
	\BibitemOpen
	\bibfield  {author} {\bibinfo {author} {\bibfnamefont {P.~M.}\ \bibnamefont
			{Sarte}}, \bibinfo {author} {\bibfnamefont {A.~M.}\ \bibnamefont
			{Ar\'evalo-L\'opez}}, \bibinfo {author} {\bibfnamefont {M.}~\bibnamefont
			{Songvilay}}, \bibinfo {author} {\bibfnamefont {D.}~\bibnamefont {Le}},
		\bibinfo {author} {\bibfnamefont {T.}~\bibnamefont {Guidi}}, \bibinfo
		{author} {\bibfnamefont {V.}~\bibnamefont {Garc\'{\i}a-Sakai}}, \bibinfo
		{author} {\bibfnamefont {S.}~\bibnamefont {Mukhopadhyay}}, \bibinfo {author}
		{\bibfnamefont {S.~C.}\ \bibnamefont {Capelli}}, \bibinfo {author}
		{\bibfnamefont {W.~D.}\ \bibnamefont {Ratcliff}}, \bibinfo {author}
		{\bibfnamefont {K.~H.}\ \bibnamefont {Hong}}, \bibinfo {author}
		{\bibfnamefont {G.~M.}\ \bibnamefont {McNally}}, \bibinfo {author}
		{\bibfnamefont {E.}~\bibnamefont {Pachoud}}, \bibinfo {author} {\bibfnamefont
			{J.~P.}\ \bibnamefont {Attfield}},\ and\ \bibinfo {author} {\bibfnamefont
			{C.}~\bibnamefont {Stock}},\ }\bibfield  {title} {\bibinfo {title} {Ordered
			magnetism in the intrinsically decorated ${j}_{\mathrm{eff}}=\frac{1}{2}$
			$\alpha$-\textsc{C}o\textsc{V}$_{3}$\textsc{O}$_{8}$},\ }\href
	{https://doi.org/10.1103/PhysRevB.98.224410} {\bibfield  {journal} {\bibinfo
			{journal} {Phys. Rev. B}\ }\textbf {\bibinfo {volume} {98}},\ \bibinfo
		{pages} {224410} (\bibinfo {year} {2018}{\natexlab{b}})}\BibitemShut
	{NoStop}%
	\bibitem [{\citenamefont {Songvilay}\ \emph {et~al.}(2020)\citenamefont
		{Songvilay}, \citenamefont {Robert}, \citenamefont {Petit}, \citenamefont
		{Rodriguez-Rivera}, \citenamefont {Ratcliff}, \citenamefont {Damay},
		\citenamefont {Bal\'edent}, \citenamefont {Jim\'enez-Ruiz}, \citenamefont
		{Lejay}, \citenamefont {Pachoud}, \citenamefont {Hadj-Azzem}, \citenamefont
		{Simonet},\ and\ \citenamefont {Stock}}]{Songvilay20:102}%
	\BibitemOpen
	\bibfield  {author} {\bibinfo {author} {\bibfnamefont {M.}~\bibnamefont
			{Songvilay}}, \bibinfo {author} {\bibfnamefont {J.}~\bibnamefont {Robert}},
		\bibinfo {author} {\bibfnamefont {S.}~\bibnamefont {Petit}}, \bibinfo
		{author} {\bibfnamefont {J.~A.}\ \bibnamefont {Rodriguez-Rivera}}, \bibinfo
		{author} {\bibfnamefont {W.~D.}\ \bibnamefont {Ratcliff}}, \bibinfo {author}
		{\bibfnamefont {F.}~\bibnamefont {Damay}}, \bibinfo {author} {\bibfnamefont
			{V.}~\bibnamefont {Bal\'edent}}, \bibinfo {author} {\bibfnamefont
			{M.}~\bibnamefont {Jim\'enez-Ruiz}}, \bibinfo {author} {\bibfnamefont
			{P.}~\bibnamefont {Lejay}}, \bibinfo {author} {\bibfnamefont
			{E.}~\bibnamefont {Pachoud}}, \bibinfo {author} {\bibfnamefont
			{A.}~\bibnamefont {Hadj-Azzem}}, \bibinfo {author} {\bibfnamefont
			{V.}~\bibnamefont {Simonet}},\ and\ \bibinfo {author} {\bibfnamefont
			{C.}~\bibnamefont {Stock}},\ }\bibfield  {title} {\bibinfo {title} {Kitaev
			interactions in the co honeycomb antiferromagnets
			\textsc{N}a$_3$\textsc{C}o$_2$\textsc{S}b\textsc{O}$_6$ and
			\textsc{N}a$_2$\textsc{C}o$_2$\textsc{T}e\textsc{O}$_6$},\ }\href
	{https://doi.org/10.1103/PhysRevB.102.224429} {\bibfield  {journal} {\bibinfo
			{journal} {Phys. Rev. B}\ }\textbf {\bibinfo {volume} {102}},\ \bibinfo
		{pages} {224429} (\bibinfo {year} {2020})}\BibitemShut {NoStop}%
	\bibitem [{\citenamefont {Stock}\ \emph {et~al.}(2009)\citenamefont {Stock},
		\citenamefont {Chapon}, \citenamefont {Adamopoulos}, \citenamefont {Lappas},
		\citenamefont {Giot}, \citenamefont {Taylor}, \citenamefont {Green},
		\citenamefont {Brown},\ and\ \citenamefont {Radaelli}}]{Stock09:103}%
	\BibitemOpen
	\bibfield  {author} {\bibinfo {author} {\bibfnamefont {C.}~\bibnamefont
			{Stock}}, \bibinfo {author} {\bibfnamefont {L.~C.}\ \bibnamefont {Chapon}},
		\bibinfo {author} {\bibfnamefont {O.}~\bibnamefont {Adamopoulos}}, \bibinfo
		{author} {\bibfnamefont {A.}~\bibnamefont {Lappas}}, \bibinfo {author}
		{\bibfnamefont {M.}~\bibnamefont {Giot}}, \bibinfo {author} {\bibfnamefont
			{J.~W.}\ \bibnamefont {Taylor}}, \bibinfo {author} {\bibfnamefont {M.~A.}\
			\bibnamefont {Green}}, \bibinfo {author} {\bibfnamefont {C.~M.}\ \bibnamefont
			{Brown}},\ and\ \bibinfo {author} {\bibfnamefont {P.~G.}\ \bibnamefont
			{Radaelli}},\ }\bibfield  {title} {\bibinfo {title} {One-dimensional magnetic
			fluctuations in the spin-2 triangular lattice
			$\alpha$-\textsc{N}a\textsc{M}n\textsc{O}$_{2}$},\ }\href
	{https://doi.org/10.1103/PhysRevLett.103.077202} {\bibfield  {journal}
		{\bibinfo  {journal} {Phys. Rev. Lett.}\ }\textbf {\bibinfo {volume} {103}},\
		\bibinfo {pages} {077202} (\bibinfo {year} {2009})}\BibitemShut {NoStop}%
	\bibitem [{\citenamefont {Kirschner}\ \emph {et~al.}(2019)\citenamefont
		{Kirschner}, \citenamefont {Johnson}, \citenamefont {Lang}, \citenamefont
		{Khalyavin}, \citenamefont {Manuel}, \citenamefont {Lancaster}, \citenamefont
		{Prabhakaran},\ and\ \citenamefont {Blundell}}]{Kirschner19:99}%
	\BibitemOpen
	\bibfield  {author} {\bibinfo {author} {\bibfnamefont {F.~K.~K.}\
			\bibnamefont {Kirschner}}, \bibinfo {author} {\bibfnamefont {R.~D.}\
			\bibnamefont {Johnson}}, \bibinfo {author} {\bibfnamefont {F.}~\bibnamefont
			{Lang}}, \bibinfo {author} {\bibfnamefont {D.~D.}\ \bibnamefont {Khalyavin}},
		\bibinfo {author} {\bibfnamefont {P.}~\bibnamefont {Manuel}}, \bibinfo
		{author} {\bibfnamefont {T.}~\bibnamefont {Lancaster}}, \bibinfo {author}
		{\bibfnamefont {D.}~\bibnamefont {Prabhakaran}},\ and\ \bibinfo {author}
		{\bibfnamefont {S.~J.}\ \bibnamefont {Blundell}},\ }\bibfield  {title}
	{\bibinfo {title} {Spin \textsc{J}ahn-\textsc{T}eller antiferromagnetism in
			\textsc{C}o\textsc{T}i$_{2}$\textsc{O}$_{5}$},\ }\href
	{https://doi.org/10.1103/PhysRevB.99.064403} {\bibfield  {journal} {\bibinfo
			{journal} {Phys. Rev. B}\ }\textbf {\bibinfo {volume} {99}},\ \bibinfo
		{pages} {064403} (\bibinfo {year} {2019})}\BibitemShut {NoStop}%
	\bibitem [{\citenamefont {Lang}\ \emph {et~al.}(2019)\citenamefont {Lang},
		\citenamefont {Jowitt}, \citenamefont {Prabhakaran}, \citenamefont
		{Johnson},\ and\ \citenamefont {Blundell}}]{Lang19:100}%
	\BibitemOpen
	\bibfield  {author} {\bibinfo {author} {\bibfnamefont {F.}~\bibnamefont
			{Lang}}, \bibinfo {author} {\bibfnamefont {L.}~\bibnamefont {Jowitt}},
		\bibinfo {author} {\bibfnamefont {D.}~\bibnamefont {Prabhakaran}}, \bibinfo
		{author} {\bibfnamefont {R.~D.}\ \bibnamefont {Johnson}},\ and\ \bibinfo
		{author} {\bibfnamefont {S.~J.}\ \bibnamefont {Blundell}},\ }\bibfield
	{title} {\bibinfo {title} {\textsc{F}e\textsc{T}i$_{2}$\textsc{O}$_{5}$: A
			spin \textsc{J}ahn-\textsc{T}eller transition enhanced by cation
			substitution},\ }\href {https://doi.org/10.1103/PhysRevB.100.094401}
	{\bibfield  {journal} {\bibinfo  {journal} {Phys. Rev. B}\ }\textbf {\bibinfo
			{volume} {100}},\ \bibinfo {pages} {094401} (\bibinfo {year}
		{2019})}\BibitemShut {NoStop}%
	\bibitem [{\citenamefont {Petricek}\ \emph {et~al.}(2014)\citenamefont
		{Petricek}, \citenamefont {Dusek},\ and\ \citenamefont
		{Palatinus}}]{Petricek14:229}%
	\BibitemOpen
	\bibfield  {author} {\bibinfo {author} {\bibfnamefont {V.}~\bibnamefont
			{Petricek}}, \bibinfo {author} {\bibfnamefont {M.}~\bibnamefont {Dusek}},\
		and\ \bibinfo {author} {\bibfnamefont {L.}~\bibnamefont {Palatinus}},\
	}\bibfield  {title} {\bibinfo {title} {Crystallographic computing system
			\textsc{JANA}2006: General features},\ }\href
	{https://doi.org/10.1515/zkri-2014-1737} {\bibfield  {journal} {\bibinfo
			{journal} {Z. Kristallogr.}\ }\textbf {\bibinfo {volume} {229(5)}},\ \bibinfo
		{pages} {345} (\bibinfo {year} {2014})}\BibitemShut {NoStop}%
	\bibitem [{\citenamefont {Coelho}(2012)}]{TOPAS}%
	\BibitemOpen
	\bibfield  {author} {\bibinfo {author} {\bibfnamefont {A.~A.}\ \bibnamefont
			{Coelho}},\ }\href@noop {} {\emph {\bibinfo {title} {TopasAcademic: General
				Profile and Structure Analysis Software for Powder Diffraction Data}}}\
	(\bibinfo  {publisher} {Bruker AXS},\ \bibinfo {address} {Karlsruhe,
		Germany},\ \bibinfo {year} {2012})\BibitemShut {NoStop}%
	\bibitem [{\citenamefont {Xu}\ \emph {et~al.}(2024)\citenamefont {Xu},
		\citenamefont {Liu}, \citenamefont {Xin}, \citenamefont {Shen}, \citenamefont
		{Luo}, \citenamefont {Zhou}, \citenamefont {Cheng}, \citenamefont {Liu},
		\citenamefont {Tao}, \citenamefont {Liu}, \citenamefont {Huo}, \citenamefont
		{Wang},\ and\ \citenamefont {Sui}}]{Xu24:33}%
	\BibitemOpen
	\bibfield  {author} {\bibinfo {author} {\bibfnamefont {H.-H.}\ \bibnamefont
			{Xu}}, \bibinfo {author} {\bibfnamefont {Q.-Y.}\ \bibnamefont {Liu}},
		\bibinfo {author} {\bibfnamefont {C.}~\bibnamefont {Xin}}, \bibinfo {author}
		{\bibfnamefont {Q.-X.}\ \bibnamefont {Shen}}, \bibinfo {author}
		{\bibfnamefont {J.}~\bibnamefont {Luo}}, \bibinfo {author} {\bibfnamefont
			{R.}~\bibnamefont {Zhou}}, \bibinfo {author} {\bibfnamefont {J.-G.}\
			\bibnamefont {Cheng}}, \bibinfo {author} {\bibfnamefont {J.}~\bibnamefont
			{Liu}}, \bibinfo {author} {\bibfnamefont {L.~L.}\ \bibnamefont {Tao}},
		\bibinfo {author} {\bibfnamefont {Z.-G.}\ \bibnamefont {Liu}}, \bibinfo
		{author} {\bibfnamefont {M.-X.}\ \bibnamefont {Huo}}, \bibinfo {author}
		{\bibfnamefont {X.-J.}\ \bibnamefont {Wang}},\ and\ \bibinfo {author}
		{\bibfnamefont {Y.}~\bibnamefont {Sui}},\ }\bibfield  {title} {\bibinfo
		{title} {Spin gap in quasi-one-dimensional $\textsc{S}=3/2$
			antiferromagnet},\ }\href {https://doi.org/10.1088/1674-1056/ad1381}
	{\bibfield  {journal} {\bibinfo  {journal} {Chin. Phys. B}\ }\textbf
		{\bibinfo {volume} {33}},\ \bibinfo {pages} {037505} (\bibinfo {year}
		{2024})}\BibitemShut {NoStop}%
	\bibitem [{\citenamefont {Wu}\ \emph {et~al.}(2021)\citenamefont {Wu},
		\citenamefont {Song}, \citenamefont {He}, \citenamefont {Frano},
		\citenamefont {Yi}, \citenamefont {Chen}, \citenamefont {Uchiyama},
		\citenamefont {Alatas}, \citenamefont {Said}, \citenamefont {Wang},
		\citenamefont {Wolf}, \citenamefont {Meingast},\ and\ \citenamefont
		{Birgeneau}}]{Wu21:126}%
	\BibitemOpen
	\bibfield  {author} {\bibinfo {author} {\bibfnamefont {S.}~\bibnamefont
			{Wu}}, \bibinfo {author} {\bibfnamefont {Y.}~\bibnamefont {Song}}, \bibinfo
		{author} {\bibfnamefont {Y.}~\bibnamefont {He}}, \bibinfo {author}
		{\bibfnamefont {A.}~\bibnamefont {Frano}}, \bibinfo {author} {\bibfnamefont
			{M.}~\bibnamefont {Yi}}, \bibinfo {author} {\bibfnamefont {X.}~\bibnamefont
			{Chen}}, \bibinfo {author} {\bibfnamefont {H.}~\bibnamefont {Uchiyama}},
		\bibinfo {author} {\bibfnamefont {A.}~\bibnamefont {Alatas}}, \bibinfo
		{author} {\bibfnamefont {A.~H.}\ \bibnamefont {Said}}, \bibinfo {author}
		{\bibfnamefont {L.}~\bibnamefont {Wang}}, \bibinfo {author} {\bibfnamefont
			{T.}~\bibnamefont {Wolf}}, \bibinfo {author} {\bibfnamefont {C.}~\bibnamefont
			{Meingast}},\ and\ \bibinfo {author} {\bibfnamefont {R.~J.}\ \bibnamefont
			{Birgeneau}},\ }\bibfield  {title} {\bibinfo {title} {Short-range nematic
			fluctuations in
			\textsc{S}r$_{1-x}$\textsc{N}a$_x$\textsc{F}e$_2$\textsc{A}s$_2$
			superconductors},\ }\href {https://doi.org/10.1103/PhysRevLett.126.107001}
	{\bibfield  {journal} {\bibinfo  {journal} {Phys. Rev. Lett.}\ }\textbf
		{\bibinfo {volume} {126}},\ \bibinfo {pages} {107001} (\bibinfo {year}
		{2021})}\BibitemShut {NoStop}%
	\bibitem [{\citenamefont {Kauth}\ \emph {et~al.}(2020)\citenamefont {Kauth},
		\citenamefont {Rosenkranz}, \citenamefont {Said}, \citenamefont {Taddei},
		\citenamefont {Wolf},\ and\ \citenamefont {Weber}}]{Kauth20:102}%
	\BibitemOpen
	\bibfield  {author} {\bibinfo {author} {\bibfnamefont {M.}~\bibnamefont
			{Kauth}}, \bibinfo {author} {\bibfnamefont {S.}~\bibnamefont {Rosenkranz}},
		\bibinfo {author} {\bibfnamefont {A.~H.}\ \bibnamefont {Said}}, \bibinfo
		{author} {\bibfnamefont {K.~M.}\ \bibnamefont {Taddei}}, \bibinfo {author}
		{\bibfnamefont {T.}~\bibnamefont {Wolf}},\ and\ \bibinfo {author}
		{\bibfnamefont {F.}~\bibnamefont {Weber}},\ }\bibfield  {title} {\bibinfo
		{title} {Soft elastic constants from phonon spectroscopy in hole-doped
			\textsc{B}a$_{1-x}$(\textsc{K},\textsc{N}a)$_x$\textsc{F}e$_2$\textsc{A}s$_2$
			and \textsc{S}r$_{1-x}$\textsc{N}a$_x$\textsc{F}e$_2$\textsc{A}s$_2$},\
	}\href {https://doi.org/10.1103/PhysRevB.102.144526} {\bibfield  {journal}
		{\bibinfo  {journal} {Phys. Rev. B}\ }\textbf {\bibinfo {volume} {102}},\
		\bibinfo {pages} {144526} (\bibinfo {year} {2020})}\BibitemShut {NoStop}%
	\bibitem [{\citenamefont {Loewenhaupt}\ \emph {et~al.}(1979)\citenamefont
		{Loewenhaupt}, \citenamefont {Rainford},\ and\ \citenamefont
		{Steglich}}]{Lowenhaupt79:42}%
	\BibitemOpen
	\bibfield  {author} {\bibinfo {author} {\bibfnamefont {M.}~\bibnamefont
			{Loewenhaupt}}, \bibinfo {author} {\bibfnamefont {B.~D.}\ \bibnamefont
			{Rainford}},\ and\ \bibinfo {author} {\bibfnamefont {F.}~\bibnamefont
			{Steglich}},\ }\bibfield  {title} {\bibinfo {title} {Dynamic
			\textsc{J}ahn-\textsc{T}eller effect in a rare-earth compound:
			\textsc{C}e\textsc{A}l$_{2}$},\ }\href
	{https://doi.org/10.1103/PhysRevLett.42.1709} {\bibfield  {journal} {\bibinfo
			{journal} {Phys. Rev. Lett.}\ }\textbf {\bibinfo {volume} {42}},\ \bibinfo
		{pages} {1709} (\bibinfo {year} {1979})}\BibitemShut {NoStop}%
	\bibitem [{\citenamefont {G\"untherodt}\ \emph {et~al.}(1983)\citenamefont
		{G\"untherodt}, \citenamefont {Jayaraman}, \citenamefont {Batlogg},
		\citenamefont {Croft},\ and\ \citenamefont {Melczer}}]{Guntherodt83:51}%
	\BibitemOpen
	\bibfield  {author} {\bibinfo {author} {\bibfnamefont {G.}~\bibnamefont
			{G\"untherodt}}, \bibinfo {author} {\bibfnamefont {A.}~\bibnamefont
			{Jayaraman}}, \bibinfo {author} {\bibfnamefont {G.}~\bibnamefont {Batlogg}},
		\bibinfo {author} {\bibfnamefont {M.}~\bibnamefont {Croft}},\ and\ \bibinfo
		{author} {\bibfnamefont {E.}~\bibnamefont {Melczer}},\ }\bibfield  {title}
	{\bibinfo {title} {Raman scattering from coupled phonon and electronic
			crystal-field excitations in \textsc{C}e\textsc{A}l$_{2}$},\ }\href
	{https://doi.org/10.1103/PhysRevLett.51.2330} {\bibfield  {journal} {\bibinfo
			{journal} {Phys. Rev. Lett.}\ }\textbf {\bibinfo {volume} {51}},\ \bibinfo
		{pages} {2330} (\bibinfo {year} {1983})}\BibitemShut {NoStop}%
	\bibitem [{\citenamefont {Thalmeier}\ and\ \citenamefont
		{Fulde}(1982)}]{Thalmeier82:49}%
	\BibitemOpen
	\bibfield  {author} {\bibinfo {author} {\bibfnamefont {P.}~\bibnamefont
			{Thalmeier}}\ and\ \bibinfo {author} {\bibfnamefont {P.}~\bibnamefont
			{Fulde}},\ }\bibfield  {title} {\bibinfo {title} {Bound state between a
			crystal-field excitation and a phonon in \textsc{C}e\textsc{A}l$_{2}$},\
	}\href {https://doi.org/10.1103/PhysRevLett.49.1588} {\bibfield  {journal}
		{\bibinfo  {journal} {Phys. Rev. Lett.}\ }\textbf {\bibinfo {volume} {49}},\
		\bibinfo {pages} {1588} (\bibinfo {year} {1982})}\BibitemShut {NoStop}%
	\bibitem [{\citenamefont {Melcher}\ and\ \citenamefont
		{Scott}(1972)}]{Melcher72:28}%
	\BibitemOpen
	\bibfield  {author} {\bibinfo {author} {\bibfnamefont {R.~L.}\ \bibnamefont
			{Melcher}}\ and\ \bibinfo {author} {\bibfnamefont {B.~A.}\ \bibnamefont
			{Scott}},\ }\bibfield  {title} {\bibinfo {title} {Soft acoustic mode at the
			cooperative \textsc{J}ahn-\textsc{T}eller phase transition in
			\textsc{D}y\textsc{V}\textsc{O}$_4$},\ }\href
	{https://doi.org/10.1103/PhysRevLett.28.607} {\bibfield  {journal} {\bibinfo
			{journal} {Phys. Rev. Lett.}\ }\textbf {\bibinfo {volume} {28}},\ \bibinfo
		{pages} {607} (\bibinfo {year} {1972})}\BibitemShut {NoStop}%
	\bibitem [{\citenamefont {Pytte}(1971)}]{Pytte71:3}%
	\BibitemOpen
	\bibfield  {author} {\bibinfo {author} {\bibfnamefont {E.}~\bibnamefont
			{Pytte}},\ }\bibfield  {title} {\bibinfo {title} {Structural phase
			transitions in spinels induced by the \textsc{J}ahn-\textsc{T}eller effect},\
	}\href {https://doi.org/10.1103/PhysRevB.3.3503} {\bibfield  {journal}
		{\bibinfo  {journal} {Phys. Rev. B}\ }\textbf {\bibinfo {volume} {3}},\
		\bibinfo {pages} {3503} (\bibinfo {year} {1971})}\BibitemShut {NoStop}%
	\bibitem [{\citenamefont {Pytte}(1973)}]{Pytte73:8}%
	\BibitemOpen
	\bibfield  {author} {\bibinfo {author} {\bibfnamefont {E.}~\bibnamefont
			{Pytte}},\ }\bibfield  {title} {\bibinfo {title} {Dynamics of
			\textsc{J}ahn-\textsc{T}eller phase transitions},\ }\href
	{https://doi.org/10.1103/PhysRevB.8.3954} {\bibfield  {journal} {\bibinfo
			{journal} {Phys. Rev. B}\ }\textbf {\bibinfo {volume} {8}},\ \bibinfo {pages}
		{3954} (\bibinfo {year} {1973})}\BibitemShut {NoStop}%
	\bibitem [{\citenamefont {Liu}\ and\ \citenamefont
		{Khaliullin}(2019)}]{Liu19:122}%
	\BibitemOpen
	\bibfield  {author} {\bibinfo {author} {\bibfnamefont {H.}~\bibnamefont
			{Liu}}\ and\ \bibinfo {author} {\bibfnamefont {G.}~\bibnamefont
			{Khaliullin}},\ }\bibfield  {title} {\bibinfo {title}
		{Pseudo-\textsc{J}ahn-\textsc{T}eller effect and magnetoelastic coupling in
			spin-orbit mott insulators},\ }\href
	{https://doi.org/10.1103/PhysRevLett.122.057203} {\bibfield  {journal}
		{\bibinfo  {journal} {Phys. Rev. Lett.}\ }\textbf {\bibinfo {volume} {122}},\
		\bibinfo {pages} {057203} (\bibinfo {year} {2019})}\BibitemShut {NoStop}%
	\bibitem [{\citenamefont {Weber}\ \emph {et~al.}(2017)\citenamefont {Weber},
		\citenamefont {Roessli}, \citenamefont {Stock}, \citenamefont {Keller},
		\citenamefont {Schmalzl}, \citenamefont {Bourdarot}, \citenamefont {Georgii},
		\citenamefont {Ewings}, \citenamefont {Perry},\ and\ \citenamefont
		{B\"oni}}]{Weber17:96}%
	\BibitemOpen
	\bibfield  {author} {\bibinfo {author} {\bibfnamefont {T.}~\bibnamefont
			{Weber}}, \bibinfo {author} {\bibfnamefont {B.}~\bibnamefont {Roessli}},
		\bibinfo {author} {\bibfnamefont {C.}~\bibnamefont {Stock}}, \bibinfo
		{author} {\bibfnamefont {T.}~\bibnamefont {Keller}}, \bibinfo {author}
		{\bibfnamefont {K.}~\bibnamefont {Schmalzl}}, \bibinfo {author}
		{\bibfnamefont {F.}~\bibnamefont {Bourdarot}}, \bibinfo {author}
		{\bibfnamefont {R.}~\bibnamefont {Georgii}}, \bibinfo {author} {\bibfnamefont
			{R.~A.}\ \bibnamefont {Ewings}}, \bibinfo {author} {\bibfnamefont {R.~S.}\
			\bibnamefont {Perry}},\ and\ \bibinfo {author} {\bibfnamefont
			{P.}~\bibnamefont {B\"oni}},\ }\bibfield  {title} {\bibinfo {title}
		{Transverse acoustic phonon anomalies at intermediate wave vectors in
			\textsc{M}g\textsc{V}$_2$\textsc{O}$_4$},\ }\href
	{https://doi.org/10.1103/PhysRevB.96.184301} {\bibfield  {journal} {\bibinfo
			{journal} {Phys. Rev. B}\ }\textbf {\bibinfo {volume} {96}},\ \bibinfo
		{pages} {184301} (\bibinfo {year} {2017})}\BibitemShut {NoStop}%
	\bibitem [{\citenamefont {Lane}\ \emph {et~al.}(2023)\citenamefont {Lane},
		\citenamefont {Sarte}, \citenamefont {Guratinder}, \citenamefont
		{Arevalo-Lopez}, \citenamefont {Perry}, \citenamefont {Hunter}, \citenamefont
		{Weber}, \citenamefont {Roessli}, \citenamefont {Stunault}, \citenamefont
		{Su}, \citenamefont {Ewings}, \citenamefont {Wilson}, \citenamefont {B\"oni},
		\citenamefont {Attfield},\ and\ \citenamefont {Stock}}]{Lane23:5}%
	\BibitemOpen
	\bibfield  {author} {\bibinfo {author} {\bibfnamefont {H.}~\bibnamefont
			{Lane}}, \bibinfo {author} {\bibfnamefont {P.~M.}\ \bibnamefont {Sarte}},
		\bibinfo {author} {\bibfnamefont {K.}~\bibnamefont {Guratinder}}, \bibinfo
		{author} {\bibfnamefont {A.~M.}\ \bibnamefont {Arevalo-Lopez}}, \bibinfo
		{author} {\bibfnamefont {R.~S.}\ \bibnamefont {Perry}}, \bibinfo {author}
		{\bibfnamefont {E.~C.}\ \bibnamefont {Hunter}}, \bibinfo {author}
		{\bibfnamefont {T.}~\bibnamefont {Weber}}, \bibinfo {author} {\bibfnamefont
			{B.}~\bibnamefont {Roessli}}, \bibinfo {author} {\bibfnamefont
			{A.}~\bibnamefont {Stunault}}, \bibinfo {author} {\bibfnamefont
			{Y.}~\bibnamefont {Su}}, \bibinfo {author} {\bibfnamefont {R.~A.}\
			\bibnamefont {Ewings}}, \bibinfo {author} {\bibfnamefont {S.~D.}\
			\bibnamefont {Wilson}}, \bibinfo {author} {\bibfnamefont {P.}~\bibnamefont
			{B\"oni}}, \bibinfo {author} {\bibfnamefont {J.~P.}\ \bibnamefont
			{Attfield}},\ and\ \bibinfo {author} {\bibfnamefont {C.}~\bibnamefont
			{Stock}},\ }\bibfield  {title} {\bibinfo {title} {Spin-orbital correlations
			from complex orbital order in \textsc{M}g\textsc{V}$_2$\textsc{O}$_4$},\
	}\href {https://doi.org/10.1103/PhysRevResearch.5.043146} {\bibfield
		{journal} {\bibinfo  {journal} {Phys. Rev. Res.}\ }\textbf {\bibinfo {volume}
			{5}},\ \bibinfo {pages} {043146} (\bibinfo {year} {2023})}\BibitemShut
	{NoStop}%
	\bibitem [{\citenamefont {Birgeneau}\ \emph {et~al.}(1974)\citenamefont
		{Birgeneau}, \citenamefont {Kjems}, \citenamefont {Shirane},\ and\
		\citenamefont {Van~Uitert}}]{Birgeneau74:10}%
	\BibitemOpen
	\bibfield  {author} {\bibinfo {author} {\bibfnamefont {R.~J.}\ \bibnamefont
			{Birgeneau}}, \bibinfo {author} {\bibfnamefont {J.~K.}\ \bibnamefont
			{Kjems}}, \bibinfo {author} {\bibfnamefont {G.}~\bibnamefont {Shirane}},\
		and\ \bibinfo {author} {\bibfnamefont {L.~G.}\ \bibnamefont {Van~Uitert}},\
	}\bibfield  {title} {\bibinfo {title} {Cooperative
			\textsc{J}ahn-\textsc{T}eller phase transition in
			\textsc{P}r\textsc{A}l\textsc{O}$_3$},\ }\href
	{https://doi.org/10.1103/PhysRevB.10.2512} {\bibfield  {journal} {\bibinfo
			{journal} {Phys. Rev. B}\ }\textbf {\bibinfo {volume} {10}},\ \bibinfo
		{pages} {2512} (\bibinfo {year} {1974})}\BibitemShut {NoStop}%
	\bibitem [{\citenamefont {Cowley}(1976)}]{Cowley76:13}%
	\BibitemOpen
	\bibfield  {author} {\bibinfo {author} {\bibfnamefont {R.~A.}\ \bibnamefont
			{Cowley}},\ }\bibfield  {title} {\bibinfo {title} {Acoustic phonon
			instabilities and structural phase transitions},\ }\href
	{https://doi.org/10.1103/PhysRevB.13.4877} {\bibfield  {journal} {\bibinfo
			{journal} {Phys. Rev. B}\ }\textbf {\bibinfo {volume} {13}},\ \bibinfo
		{pages} {4877} (\bibinfo {year} {1976})}\BibitemShut {NoStop}%
	\bibitem [{\citenamefont {Dove}(1993)}]{Dove:book}%
	\BibitemOpen
	\bibfield  {author} {\bibinfo {author} {\bibfnamefont {M.~T.}\ \bibnamefont
			{Dove}},\ }\href@noop {} {\emph {\bibinfo {title} {Introduction to Lattice
				Dynamics}}}\ (\bibinfo  {publisher} {Cambridge University Press},\ \bibinfo
	{address} {Cambridge},\ \bibinfo {year} {1993})\BibitemShut {NoStop}%
	\bibitem [{\citenamefont {Maradudin}(1962)}]{Maradudin62:2}%
	\BibitemOpen
	\bibfield  {author} {\bibinfo {author} {\bibfnamefont {A.~A.}\ \bibnamefont
			{Maradudin}},\ }\bibfield  {title} {\bibinfo {title} {Thermal expansion and
			phonon frequency shifts},\ }\href {https://doi.org/10.1002/pssb.19620021107}
	{\bibfield  {journal} {\bibinfo  {journal} {Phys. Status Solidi B}\ }\textbf
		{\bibinfo {volume} {2}},\ \bibinfo {pages} {1493} (\bibinfo {year}
		{1962})}\BibitemShut {NoStop}%
	\bibitem [{\citenamefont {Guratinder}\ \emph {et~al.}(2023)\citenamefont
		{Guratinder}, \citenamefont {Chan}, \citenamefont {Rodriguez}, \citenamefont
		{Rodriguez-Rivera}, \citenamefont {Stuhr}, \citenamefont {Stunault},
		\citenamefont {Travers}, \citenamefont {Green}, \citenamefont {Qureshi},\
		and\ \citenamefont {Stock}}]{Guratinder23:108}%
	\BibitemOpen
	\bibfield  {author} {\bibinfo {author} {\bibfnamefont {K.}~\bibnamefont
			{Guratinder}}, \bibinfo {author} {\bibfnamefont {E.}~\bibnamefont {Chan}},
		\bibinfo {author} {\bibfnamefont {E.~E.}\ \bibnamefont {Rodriguez}}, \bibinfo
		{author} {\bibfnamefont {J.~A.}\ \bibnamefont {Rodriguez-Rivera}}, \bibinfo
		{author} {\bibfnamefont {U.}~\bibnamefont {Stuhr}}, \bibinfo {author}
		{\bibfnamefont {A.}~\bibnamefont {Stunault}}, \bibinfo {author}
		{\bibfnamefont {R.}~\bibnamefont {Travers}}, \bibinfo {author} {\bibfnamefont
			{M.~A.}\ \bibnamefont {Green}}, \bibinfo {author} {\bibfnamefont
			{N.}~\bibnamefont {Qureshi}},\ and\ \bibinfo {author} {\bibfnamefont
			{C.}~\bibnamefont {Stock}},\ }\bibfield  {title} {\bibinfo {title} {Acoustic
			lattice instabilities at the magnetostructural transition in
			\textsc{F}e$_{1.057(7)}$\textsc{T}e},\ }\href
	{https://doi.org/10.1103/PhysRevB.108.214411} {\bibfield  {journal} {\bibinfo
			{journal} {Phys. Rev. B}\ }\textbf {\bibinfo {volume} {108}},\ \bibinfo
		{pages} {214411} (\bibinfo {year} {2023})}\BibitemShut {NoStop}%
	\bibitem [{\citenamefont {Dudarev}\ \emph {et~al.}(1998)\citenamefont
		{Dudarev}, \citenamefont {Botton}, \citenamefont {Savrasov}, \citenamefont
		{Humphreys},\ and\ \citenamefont {Sutton}}]{Dudarev98:57}%
	\BibitemOpen
	\bibfield  {author} {\bibinfo {author} {\bibfnamefont {S.~L.}\ \bibnamefont
			{Dudarev}}, \bibinfo {author} {\bibfnamefont {G.~A.}\ \bibnamefont {Botton}},
		\bibinfo {author} {\bibfnamefont {S.~Y.}\ \bibnamefont {Savrasov}}, \bibinfo
		{author} {\bibfnamefont {C.~J.}\ \bibnamefont {Humphreys}},\ and\ \bibinfo
		{author} {\bibfnamefont {A.~P.}\ \bibnamefont {Sutton}},\ }\bibfield  {title}
	{\bibinfo {title} {Electron-energy-loss spectra and the structural stability
			of nickel oxide: An \textsc{LSDA+U} study},\ }\href
	{https://doi.org/10.1103/PhysRevB.57.1505} {\bibfield  {journal} {\bibinfo
			{journal} {Phys. Rev. B}\ }\textbf {\bibinfo {volume} {57}},\ \bibinfo
		{pages} {1505} (\bibinfo {year} {1998})}\BibitemShut {NoStop}%
	\bibitem [{\citenamefont {Liechtenstein}\ \emph {et~al.}(1995)\citenamefont
		{Liechtenstein}, \citenamefont {Anisimov},\ and\ \citenamefont
		{Zaanen}}]{Liechtenstein95:52}%
	\BibitemOpen
	\bibfield  {author} {\bibinfo {author} {\bibfnamefont {A.~I.}\ \bibnamefont
			{Liechtenstein}}, \bibinfo {author} {\bibfnamefont {V.~I.}\ \bibnamefont
			{Anisimov}},\ and\ \bibinfo {author} {\bibfnamefont {J.}~\bibnamefont
			{Zaanen}},\ }\bibfield  {title} {\bibinfo {title} {Density-functional theory
			and strong interactions: Orbital ordering in mott-hubbard insulators},\
	}\href {https://doi.org/10.1103/PhysRevB.52.R5467} {\bibfield  {journal}
		{\bibinfo  {journal} {Phys. Rev. B}\ }\textbf {\bibinfo {volume} {52}},\
		\bibinfo {pages} {R5467} (\bibinfo {year} {1995})}\BibitemShut {NoStop}%
	\bibitem [{\citenamefont {Maksimov}\ \emph {et~al.}(2022)\citenamefont
		{Maksimov}, \citenamefont {Ushakov}, \citenamefont {Pchelkina}, \citenamefont
		{Li}, \citenamefont {Winter},\ and\ \citenamefont
		{Streltsov}}]{Maksimov22:106}%
	\BibitemOpen
	\bibfield  {author} {\bibinfo {author} {\bibfnamefont {P.~A.}\ \bibnamefont
			{Maksimov}}, \bibinfo {author} {\bibfnamefont {A.~V.}\ \bibnamefont
			{Ushakov}}, \bibinfo {author} {\bibfnamefont {Z.~V.}\ \bibnamefont
			{Pchelkina}}, \bibinfo {author} {\bibfnamefont {Y.}~\bibnamefont {Li}},
		\bibinfo {author} {\bibfnamefont {S.~M.}\ \bibnamefont {Winter}},\ and\
		\bibinfo {author} {\bibfnamefont {S.~V.}\ \bibnamefont {Streltsov}},\
	}\bibfield  {title} {\bibinfo {title} {Ab initio guided minimal model for the
			``\textsc{K}itaev'' material
			\textsc{B}a\textsc{C}o$_2$(\textsc{A}s\textsc{O}$_4$)$_2$: Importance of
			direct hopping, third-neighbor exchange, and quantum fluctuations},\ }\href
	{https://doi.org/10.1103/PhysRevB.106.165131} {\bibfield  {journal} {\bibinfo
			{journal} {Phys. Rev. B}\ }\textbf {\bibinfo {volume} {106}},\ \bibinfo
		{pages} {165131} (\bibinfo {year} {2022})}\BibitemShut {NoStop}%
	\bibitem [{\citenamefont {Streltsov}\ \emph {et~al.}(2005)\citenamefont
		{Streltsov}, \citenamefont {Mylnikova}, \citenamefont {Shorikov},
		\citenamefont {Pchelkina}, \citenamefont {Khomskii},\ and\ \citenamefont
		{Anisimov}}]{Streltsov05:71}%
	\BibitemOpen
	\bibfield  {author} {\bibinfo {author} {\bibfnamefont {S.~V.}\ \bibnamefont
			{Streltsov}}, \bibinfo {author} {\bibfnamefont {A.~S.}\ \bibnamefont
			{Mylnikova}}, \bibinfo {author} {\bibfnamefont {A.~O.}\ \bibnamefont
			{Shorikov}}, \bibinfo {author} {\bibfnamefont {Z.~V.}\ \bibnamefont
			{Pchelkina}}, \bibinfo {author} {\bibfnamefont {D.~I.}\ \bibnamefont
			{Khomskii}},\ and\ \bibinfo {author} {\bibfnamefont {V.~I.}\ \bibnamefont
			{Anisimov}},\ }\bibfield  {title} {\bibinfo {title} {Crystal-field splitting
			for low symmetry systems in ab initio calculations},\ }\href
	{https://doi.org/10.1103/PhysRevB.71.245114} {\bibfield  {journal} {\bibinfo
			{journal} {Phys. Rev. B}\ }\textbf {\bibinfo {volume} {71}},\ \bibinfo
		{pages} {245114} (\bibinfo {year} {2005})}\BibitemShut {NoStop}%
	\bibitem [{\citenamefont {Behr}\ \emph {et~al.}(2024)\citenamefont {Behr},
		\citenamefont {Taran}, \citenamefont {Porter}, \citenamefont {Bombardi},
		\citenamefont {Prabhakaran}, \citenamefont {Streltsov},\ and\ \citenamefont
		{Johnson}}]{Behr24:110}%
	\BibitemOpen
	\bibfield  {author} {\bibinfo {author} {\bibfnamefont {D.}~\bibnamefont
			{Behr}}, \bibinfo {author} {\bibfnamefont {L.~S.}\ \bibnamefont {Taran}},
		\bibinfo {author} {\bibfnamefont {D.~G.}\ \bibnamefont {Porter}}, \bibinfo
		{author} {\bibfnamefont {A.}~\bibnamefont {Bombardi}}, \bibinfo {author}
		{\bibfnamefont {D.}~\bibnamefont {Prabhakaran}}, \bibinfo {author}
		{\bibfnamefont {S.~V.}\ \bibnamefont {Streltsov}},\ and\ \bibinfo {author}
		{\bibfnamefont {R.~D.}\ \bibnamefont {Johnson}},\ }\bibfield  {title}
	{\bibinfo {title} {Strain-induced antiferromagnetic domain switching via the
			spin \textsc{J}ahn-\textsc{T}eller effect},\ }\href
	{https://doi.org/10.1103/PhysRevB.110.L060408} {\bibfield  {journal}
		{\bibinfo  {journal} {Phys. Rev. B}\ }\textbf {\bibinfo {volume} {110}},\
		\bibinfo {pages} {L060408} (\bibinfo {year} {2024})}\BibitemShut {NoStop}%
	\bibitem [{\citenamefont {Stoudenmire}\ \emph {et~al.}(2009)\citenamefont
		{Stoudenmire}, \citenamefont {Trebst},\ and\ \citenamefont
		{Balents}}]{Stoudenmire09:79}%
	\BibitemOpen
	\bibfield  {author} {\bibinfo {author} {\bibfnamefont {E.~M.}\ \bibnamefont
			{Stoudenmire}}, \bibinfo {author} {\bibfnamefont {S.}~\bibnamefont
			{Trebst}},\ and\ \bibinfo {author} {\bibfnamefont {L.}~\bibnamefont
			{Balents}},\ }\bibfield  {title} {\bibinfo {title} {Quadrupolar correlations
			and spin freezing in $\textsc{S}=1$ triangular lattice antiferromagnets},\
	}\href {https://doi.org/10.1103/PhysRevB.79.214436} {\bibfield  {journal}
		{\bibinfo  {journal} {Phys. Rev. B}\ }\textbf {\bibinfo {volume} {79}},\
		\bibinfo {pages} {214436} (\bibinfo {year} {2009})}\BibitemShut {NoStop}%
	\bibitem [{\citenamefont {Valentine}\ \emph {et~al.}(2020)\citenamefont
		{Valentine}, \citenamefont {Higo}, \citenamefont {Nambu}, \citenamefont
		{Chaudhuri}, \citenamefont {Wen}, \citenamefont {Broholm}, \citenamefont
		{Nakatsuji},\ and\ \citenamefont {Drichko}}]{Valentine20:6}%
	\BibitemOpen
	\bibfield  {author} {\bibinfo {author} {\bibfnamefont {M.~E.}\ \bibnamefont
			{Valentine}}, \bibinfo {author} {\bibfnamefont {T.}~\bibnamefont {Higo}},
		\bibinfo {author} {\bibfnamefont {Y.}~\bibnamefont {Nambu}}, \bibinfo
		{author} {\bibfnamefont {D.}~\bibnamefont {Chaudhuri}}, \bibinfo {author}
		{\bibfnamefont {J.}~\bibnamefont {Wen}}, \bibinfo {author} {\bibfnamefont
			{C.}~\bibnamefont {Broholm}}, \bibinfo {author} {\bibfnamefont
			{S.}~\bibnamefont {Nakatsuji}},\ and\ \bibinfo {author} {\bibfnamefont
			{N.}~\bibnamefont {Drichko}},\ }\bibfield  {title} {\bibinfo {title} {Impact
			of the lattice on magnetic properties and possible spin nematicity in the
			$\textsc{S}=1$ triangular antiferromagnet
			\textsc{N}i\textsc{G}a$_2$\textsc{S}$_4$},\ }\href
	{https://doi.org/10.1103/PhysRevLett.125.197201} {\bibfield  {journal}
		{\bibinfo  {journal} {Phys. Rev. Lett.}\ }\textbf {\bibinfo {volume} {125}},\
		\bibinfo {pages} {197201} (\bibinfo {year} {2020})}\BibitemShut {NoStop}%
	\bibitem [{\citenamefont {Tsunetsugu}\ and\ \citenamefont
		{Arikawa}(2006)}]{Tsunetsugu06:75}%
	\BibitemOpen
	\bibfield  {author} {\bibinfo {author} {\bibfnamefont {H.}~\bibnamefont
			{Tsunetsugu}}\ and\ \bibinfo {author} {\bibfnamefont {M.}~\bibnamefont
			{Arikawa}},\ }\bibfield  {title} {\bibinfo {title} {Spin nematic phase in
			$\textsc{S}=1$ triangular antiferromagnetis},\ }\href
	{https://doi.org/10.1143/JPSJ.75.083701} {\bibfield  {journal} {\bibinfo
			{journal} {J. Phys. Soc. Jpn.}\ }\textbf {\bibinfo {volume} {75}},\ \bibinfo
		{pages} {083701} (\bibinfo {year} {2006})}\BibitemShut{NoStop}%
\bibitem [{\citenamefont {Anderson}\ \emph {et~al.}(2019)\citenamefont
	{Anderson}, \citenamefont {Giri}, \citenamefont {Vinci},\ and\ \citenamefont
	{Chan}}]{anderson2019}%
\BibitemOpen
\bibfield  {author} {\bibinfo {author} {\bibfnamefont {Kevin~P.}~\bibnamefont
	{Anderson}}, \bibinfo {author} {\bibfnamefont {Anit~K.}~\bibnamefont
	{Giri}}, \bibinfo {author} {\bibfnamefont {Richard~P.}~\bibnamefont
	{Vinci}},\ and\ \bibinfo {author} {\bibfnamefont {Helen~M.}~\bibnamefont
	{Chan}},\ }\bibfield  {title} {\bibinfo {title} {Single crystal growth of 
	\textsc{C}o\textsc{T}i$_{2}$\textsc{O}$_{5}$ by solid state reaction 
	synthesis},\ }\href {https://doi.org/10.1111/jace.16379} {\bibfield  
	{journal} {\bibinfo  {journal} {J. Am. Ceram. Soc.}\ }\textbf {\bibinfo 
	{volume} {102}},\ \bibinfo {pages} {5050--5062} (\bibinfo {year} 
	{2019})}\BibitemShut{NoStop}%
\end{thebibliography}

%

\end{document}


\title{Magnetoelastic dynamics of the spin Jahn-Teller transition in CoTi$_{2}$O$_{5}$}

\author{K. Guratinder}
\affiliation{School of Physics and Astronomy, University of Edinburgh, Edinburgh EH9 3JZ, UK}
\author{R. D. Johnson}
\affiliation{Department of Physics and Astronomy, University College London, Gower St., London, WC1E 6BT, UK}
\affiliation{London Centre for Nanotechnology, University College London, Gordon Street, London WC1H 0AH, UK}
\author{D. Prabhakaran}
\affiliation{Department of Physics, Clarendon Laboratory, University of Oxford, Parks Road, Oxford OX1 3PU, UK}
\author{R. A. Taylor}
\affiliation{Department of Physics, Clarendon Laboratory, University of Oxford, Parks Road, Oxford OX1 3PU, UK}
\author{F. Lang}
\affiliation{Department of Physics, Clarendon Laboratory, University of Oxford, Parks Road, Oxford OX1 3PU, UK}
\author{S. J. Blundell}
\affiliation{Department of Physics, Clarendon Laboratory, University of Oxford, Parks Road, Oxford OX1 3PU, UK}
\author{L. S. Taran}
\affiliation{M. N. Mikheev Institute of Metal Physics, Ural Branch of Russian Academy of Sciences, 620137 Yekaterinburg, Russia}
\author{S. V. Streltsov}
\affiliation{M. N. Mikheev Institute of Metal Physics, Ural Branch of Russian Academy of Sciences, 620137 Yekaterinburg, Russia}
\affiliation{Institute of Physics and Technology, Ural Federal University, 620002 Yekaterinburg, Russia}
\author{T. J. Williams}
\affiliation{ISIS Pulsed Neutron and Muon Source, STFC Rutherford Appleton Laboratory, Harwell Campus, Didcot, Oxon OX11 0QX, United Kingdom}
\author{S. R. Giblin}
\affiliation{School of Physics and Astronomy, Cardiff University, Cardiff CF24 3AA, UK}
\author{T. Fennell}
\affiliation{Laboratory for Neutron Scattering, Paul Scherrer Institut, CH-5232 Villigen, Switzerland}
\author{K. Schmalzl}
\affiliation{Julich Centre for Neutron Science, Forschungszentrum Julich GmbH, Outstation at Institut Laue-Langevin, Boite Postal 156, 38042 Grenoble Cedex 9, France}
\author{C. Stock}
\affiliation{School of Physics and Astronomy, University of Edinburgh, Edinburgh EH9 3JZ, UK}

\date{\today}

\begin{abstract}

Supplementary information is provided on experimental aspects, phonon lineshapes, magnetic dispersions indicative of coupling, and finally density functional theory compared with Raman.

\end{abstract}

\pacs{}

\maketitle

\renewcommand{\thefigure}{S1}
\begin{figure}[ht!]
\begin{center}
	\includegraphics[width=65mm,trim=6cm 6.5cm 7cm 6.0cm,clip=true]{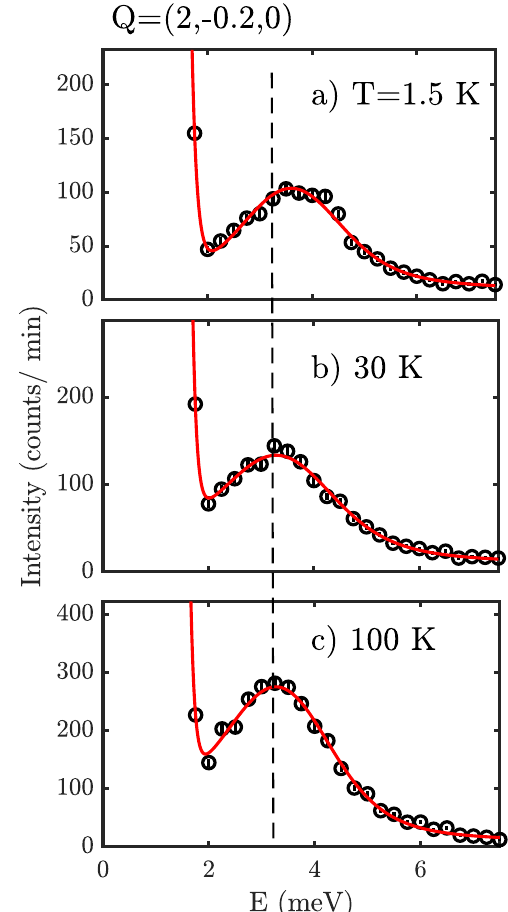}
\end{center}
\caption{Representative constant momentum scans through the acoustic phonon discussed in the main text.  The dashed line illustrates a hardening of the peak position at low temperatures with a complete temperature dependence finding abrupt onset at $T_{\rm N}$.} 
\label{fig:phonon_constQ}
\end{figure}

\section{Neutron Experiments}
Neutron experiments presented in the main text were done on two different spectrometers.  Thermal neutron measurements were performed on the Eiger triple-axis spectrometer at PSI (Switzerland) which allows specific regions in momentum and energy to be studied in detail as required for studying low-energy acoustic phonons.  Pyrolytic PG(002) reflections were used to select and incident energy ($E_i$) and a final energy ($E_f$) of 14.6 meV was selected with a PG(002) analyzer with energy transfer defined as $\hbar \omega=E_i-E_f$.  A PG filter was used on the scattered side to filter out higher order reflections.  IN12 at the ILL (France) was used to measure the low-energy magnetic excitations and check for any dispersion along the K direction to test for coupling between the buckled sheets.  PG(002) was used to both monochromate and analyze neutrons with a fixed final energy of $E_f$=5 meV.  Supplemental experiments checking for the dispersion along the L direction (orthogonal to the scattering plane used for our triple-axis experiments) and also to search for spin-orbit excitations that would indicate an orbital degeneracy were done on the MAPS time-of-flight spectrometer at ISIS (UK).  A Fermi chopper spun at a frequency of 300 Hz was used to select either incident energies of 75 or 50 meV. The single crystals used for the neutron experiments at PSI, ILL, and ISIS are shown in Fig \ref{fig:Xtal}. 

\renewcommand{\thefigure}{S2}
\begin{figure}[ht!]
\begin{center}
	\includegraphics[width=65mm,trim=0cm 0cm 0cm 0cm,clip=true]{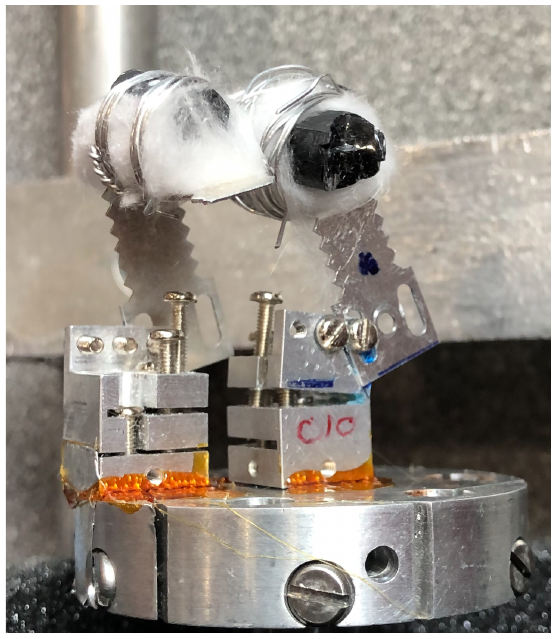}
\end{center}
\caption{The co-aligned single crystals of CoTi$_2$O$_5$ used for the neutron experiments.  One crystal was used for the IN12 and the other crystal was used for the EIGER data.  The crystals were combined for the MAPS experiment.} 
\label{fig:Xtal}
\end{figure}

\section{Raman Spectroscopy}

Raman spectroscopy measurements were performed using a microPL system. A CW 532 nm diode laser was employed with a Semrock 532 nm notch filter positioned after the sample. The light was focussed onto the sample using a 100x Mitutoyo long working distance objective with a 0.7 NA. The Raman emission was collected using the same lens and focussed onto a Princeton Instruments 0.5 m focal length spectrometer and detected using a Pylon Princeton Instruments nitrogen cooled CCD. The sample was mounted within a Janis ST-500 continuous flow optical cryostat in reflection geometry, with the crystallographic $c$ axis surface normal.

\section{Phonon lineshapes}

In this section, we outline the lineshape used to analyze the phonon data from Eiger and illustrate representative fits.  To obtain the energy position and linewidth of the acoustic phonons describe in the main text, we have fit constant-$Q$ scans to a simple harmonic oscillator being defined by the antisymmetric Lorentzian,

\begin{equation}\label{eq:SHO}
\chi''(\omega) \propto \left({1 \over {1+\left({{\omega-\Omega_{0}} \over \Gamma}\right)^{2}}}- {1 \over {1+\left({{\omega+\Omega_{0}} \over \Gamma}\right)^{2}}}\right)
\end{equation}

\noindent with $\hbar \Gamma$ the energy linewidth, inversely proportional to the lifetime $\tau$, and with an energy position of $\hbar \Omega_{0}$.  The Lorentzian at $+\hbar \Omega_0$ in Eqn. \ref{eq:SHO} accounts for the cross section for neutron energy gain (energy transfer negative $E\equiv \hbar \omega < 0$).  This is required for the cross section to follow detailed balance, which implies that the imaginary part of the susceptibility $\chi '' (\vec{Q},E)$ is an odd function in energy.  To obtain the full neutron cross-section we multiply Eqn. \ref{eq:SHO} by the thermal bose factor.

Representative fits to this lineshape are illustrated in Fig. \ref{fig:phonon_constQ} at T= 1.5 K, 30 K, and 100 K with the lineshape described above convolved with the experimental resolution obtained from scanning through the incoherent elastic line.  The phonon energy can be seen to be constant within experimental error but shifts at low temperatures to higher energies as reported in the main text.   

\renewcommand{\thefigure}{S3}
\begin{figure}[t]
\begin{center}
	\includegraphics[width=65mm,trim=5cm 5cm 5cm 5cm,clip=true]{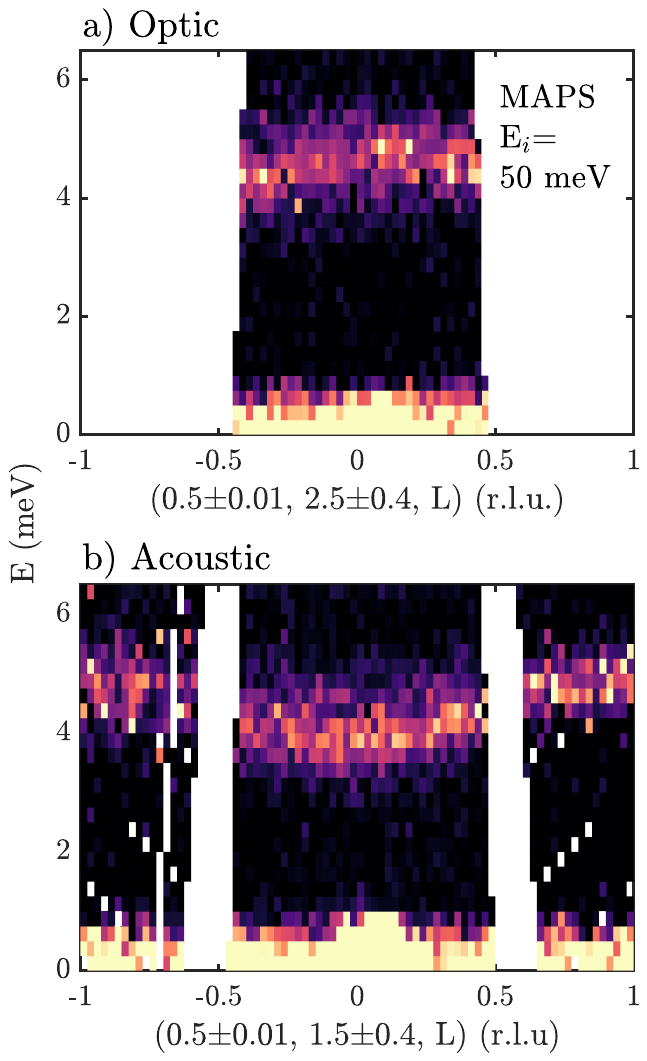}
\end{center}
\caption{$(a)$ illustrates the L dependence of the optic mode and $(b)$ the acoustic mode discussed in Figure 2 of the main text.} 
\label{fig:magnon}
\end{figure}
\section{Magnon dispersions and the question of spin-orbit coupling}

\renewcommand{\thefigure}{S4}
\begin{figure}[t]
\begin{center}
	\includegraphics[width=80mm,trim=4cm 8cm 4cm 8.5cm,clip=true]{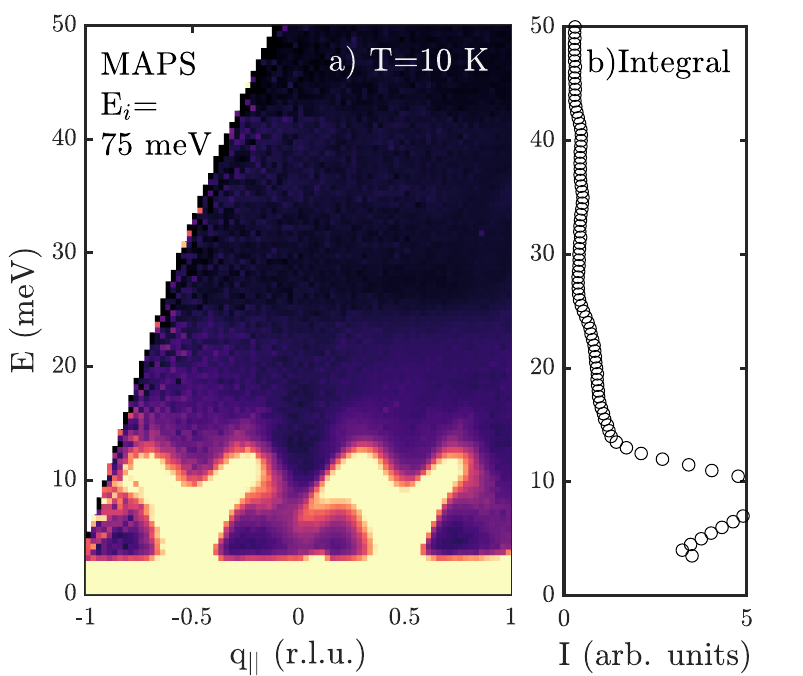}
\end{center}
\caption{A momentum-energy slice integrating along both the K and L directions given the small dispersion of the magnetic excitations along these two directions reported above and in the main text.  Not clear spin-orbit transition is observable in the range of 20-30 meV, indicative of a lack of an orbital degree of freedom.  This supports the presence of spin-only magnetism in CoTi$_2$O$_5$.} 
\label{fig:spinorbit}
\end{figure}

In this section, we discuss supplementary data relating to the magnon dispersion and also discuss the question of an orbital degree of freedom.  For this section, we apply the MAPS spectrometer from ISIS that allows measurements along all three crystallographic directions owing to the arrays of position sensitive detectors.

In Figure 2 of the main text we analyzed the magnetic scattering in the (H, K, 0) scattering plane.  In Figure 2 $(a)$ we observed two separate magnon modes with structure factors that related to optic- and acoustic- like responses.  In Fig. \ref{fig:magnon} we illustrate magnon dispersions out of the (H, K, 0) plane and along the (0,0,L) directions for both acoustic and optic modes identified in Figure 2 of the main text.  Fig. \ref{fig:magnon} $(a)$ is cut such that it integrates in K over the optic magnon mode.  In contrast Fig. \ref{fig:magnon} $(b)$ illustrates a cut along L for the acoustic magnon mode.  We note that larger out-of-plane detector coverage at lower angles facilitated a more complete measurement of the dispersion mode for the acoustic-like mode which is readily accessible at small scattering angles.  Both cuts for the acoustic- and optic-like magnons illustrate a weak dispersion along the L direction with optic and acoustic magnons being out of phase with respect to each other.  While the dispersion along L is weak, it is measurable with the resolution on MAPS.  This is in contrast to the dispersionless magnon modes along K measured with much better energy resolution on the IN12 cold neutron spectrometers.  We note that IN12 also observed a strong dispersion along the H direction.  The dispersing modes along both H and L, but not K are indicative of a weakly two-dimensional system.

From the low-energy magnetic excitations presented in Fig. \ref{fig:magnon} from MAPS and Figure 2 of the main paper taken on IN12, we can provide an estimate of the exchange constants.   Figure 2 of the main paper illustrates strongly dispersing excitations along the [1,0,0] while no dispersion is observable along [0,1,0] and only a weak dispersion observable along [0,0,1] in Fig. \ref{fig:magnon}.  This reflects a strong one-dimensional coupling along [1,0,0] and we fit the dispersion in Figure 2 of the main paper to the following result based on a magnetic Hamiltonian of,

\begin{equation}
\mathcal{H}={J \over 2} \sum_{i,j} \vec{S}_{i} \cdot \vec{S}_{j} + D \sum_{i} S_{z,i}^2
\end{equation}

\noindent for magnetic chains,

\begin{equation}
E^2(\omega) = 4S^2 \left[ D + 2JD + J^2 \sin^2 (\mathbf{Q} \cdot \mathbf{d}) \right]
\end{equation}

\noindent where $\vec{d}$ is the displacement vector connecting two nearest neighbors.  $J$ is the Heisenberg exchange coupling, $D$ is the anisotropy, and $S$ is the magnitude of the spin taken to be $S={3\over 2}$.  Based on this analysis we estimate an exchange constant of $J=3.3 \pm 0.2$ meV along [1,0,0], $0.30\pm 0.15$ meV along [0,0,1], and less than 0.07 meV along [0,1,0] with the value being constrained by the resolution of the IN12 cold triple-axis.  The single-ion anisotropy was fit to be $D=0.25 \pm 0.12$ meV.  

\begin{table}[!ht]
\centering
\begin{tabular}{|l|l|l|}
	\hline
	Mode & & peak, cm$^{-1}$ \\ \hline
	{\color{green} $A_g$}	& {\color{green} $\Gamma_{1}^{+}$} & {\color{green}746.836363}\\ \hline
	{\color{cyan} $B_{3g}$}	& {\color{cyan} $\Gamma_{3}^{+}$}  & {\color{cyan}724.872416}\\ \hline
	{\color{red} $B_{1g}$}	& {\color{red} $\Gamma_{2}^{+}$}  & {\color{red} 551.899602}\\ \hline
	{\color{blue} $B_{2g}$}	& {\color{blue} $\Gamma_{4}^{+}$} & {\color{blue}658.088127}\\ \hline
	{\color{cyan} $B_{3g}$}	& {\color{cyan} $\Gamma_{3}^{+}$} & {\color{cyan}612.862505}\\ \hline
	{\color{green} $A_g$}	& {\color{green} $\Gamma_{1}^{+}$} & {\color{green} 476.145407}\\ \hline
	{\color{cyan} $B_{3g}$}	& {\color{cyan} $\Gamma_{3}^{+}$} & {\color{cyan}439.818744}\\ \hline
	{\color{green} $A_g$}	& {\color{green} $\Gamma_{1}^{+}$} & {\color{green} 420.560745}\\ \hline
	{\color{red} $B_{1g}$}	& {\color{red} $\Gamma_{2}^{+}$} & {\color{red} 420.309003}\\ \hline
	{\color{cyan} $B_{3g}$}	& {\color{cyan} $\Gamma_{3}^{+}$} & {\color{cyan}385.202626}\\ \hline
	{\color{green} $A_g$}	& {\color{green} $\Gamma_{1}^{+}$} & {\color{green} 379.174099}\\ \hline
	{\color{cyan} $B_{3g}$}	& {\color{cyan} $\Gamma_{3}^{+}$} & {\color{cyan}332.338892}\\ \hline
	{\color{green} $A_g$}	& {\color{green} $\Gamma_{1}^{+}$} & {\color{green} 304.112528}\\ \hline
	{\color{green} $A_g$}	& {\color{green} $\Gamma_{1}^{+}$} & {\color{green} 271.037644}\\ \hline
	{\color{red} $B_{1g}$}	& {\color{red} $\Gamma_{2}^{+}$} & {\color{red} 247.04046}\\ \hline
	{\color{blue} $B_{2g}$}	& {\color{blue} $\Gamma_{4}^{+}$} & {\color{blue}245.25277}\\ \hline
	{\color{cyan} $B_{3g}$}	& {\color{cyan} $\Gamma_{3}^{+}$} & {\color{cyan}231.007093}\\ \hline
	{\color{green} $A_g$}	& {\color{green} $\Gamma_{1}^{+}$} & {\color{green} 220.009108} \\ \hline
	{\color{cyan} $B_{3g}$}	& {\color{cyan} $\Gamma_{3}^{+}$} & {\color{cyan}186.003747}\\ \hline
	{\color{red} $B_{1g}$}	& {\color{red} $\Gamma_{2}^{+}$} & {\color{red} 177.64073}\\ \hline
	{\color{blue} $B_{2g}$}	& {\color{blue} $\Gamma_{4}^{+}$} & {\color{blue}160.852467}\\ \hline
	{\color{cyan} $B_{3g}$}	& {\color{cyan} $\Gamma_{3}^{+}$} & {\color{cyan}152.923458}\\ \hline
	{\color{red} $B_{1g}$}	& {\color{red} $\Gamma_{2}^{+}$} & {\color{red} 132.371484}\\ \hline
	{\color{green} $A_g$}	& {\color{green} $\Gamma_{1}^{+}$} & {\color{green} 96.934138}\\ \hline
\end{tabular}
\caption{Frequencies at $\vec{Q}=0$ of all optical phonon modes in $\Gamma$ point as calculated by GGA+U. Note that 8 cm$^{-1}$ = 1 meV for comparison with results presented in the main paper.  The $\Gamma_{2}^{+}$ modes that break the frustrating $m_{x}$ mirror plane are highlighted in {\color{red} red}.} \label{tab1}
\end{table}

\begin{table}[ht]
\begin{tabular}{|l|l|l|}
	\hline
	{\color{green} $A_{1g}$} & {\color{green} $\Gamma_1^{+}$} & \textit{totally symmetric longitudinal} \\
	\hline {\color{red} $B_{1g}$} & {\color{red} $\Gamma_2^{+}$} & \textit{monoclinic shear distortion with unique $c$} \\
	\hline {\color{blue} $B_{2g}$} & {\color{blue} $\Gamma_4^{+}$} & \textit{monoclinic shear distortion with unique $b$} \\
	\hline {\color{cyan} $B_{3g}$} & {\color{cyan} $\Gamma_3^{+}$} & \textit{monoclinic shear distortion with unique $a$} \\ 
	\hline
\end{tabular}
\caption{Summary of notation used to describe the symmetry of the phonon modes and description of the excitation eigenvectors for acoustic modes.}\label{tab2}
\end{table}

\begin{table*}
\caption{Refined structural parameters for powdered CoTi$_2$O$_5$ at 300~K. Space group $Cmcm$, (63), $a$~=~3.728798(18)\AA,$b$~=~9.723612(51)\AA,$c$~=~10.082656(1)\AA, $U_{iso}$ for oxygen atoms(fixed)~=~0.01~\AA$^{2}$, goodness of fit (GOF)~=~1.4, R$_{p}$~=~7.66\, wR$_{p}$~=~10.47. Sites are assumed to be fully occupied.}
\label{T_1}
\setlength{\tabcolsep}{8pt}
\begin{tabular}{cccccc}
	\hline
	Atom & Site &    $x$     &    $y$     &     $z$    &      $U_{iso}$  \\ \hline
	Co  &  4$c$  &    0     &    0.19158(16)     & 1/4 &  0.01803(6)\\
	Ti  &  8$f$  &    0     &    0.13373(13)     & 0.56614(12)& 0.00912(4) \\
	O1  &  4$c$  &    0     &    0.77906(56)     & 1/4  &  0.01 \\
	O2  &  8$f$  &    0     &    0.04374(4)    & 0.11050(32) &   0.01 \\
	O3  &  8$f$  &    0     &    0.31974(33)     & 0.05854(33) &    0.01 \\ \hline
	
\end{tabular}

\end{table*}

\begin{table*}
\caption{Refined structural parameters for powdered CoTi$_2$O$_5$ at 12~K. Space group $Cmcm$, (63), $a$~=~3.729444(18)\AA,$b$~=~9.707705(53)\AA,$c$~=~10.067003(52)\AA, $U_{iso}$ for oxygen atoms(fixed)~=~0.01~\AA$^{2}$, goodness of fit (GOF)~=~1.39, R$_{p}$~=~7.39\, wR$_{p}$~=~10.19. Sites are assumed to be fully occupied.}
\label{T_1}
\setlength{\tabcolsep}{8pt}
\begin{tabular}{cccccc}
	\hline
	Atom & Site &    $x$     &    $y$     &     $z$    &        $U_{iso}$  \\ \hline
	Co  &  4$c$  &    0     &    0.19266(15)     & 1/4 &     0.01572(5)\\
	Ti  &  8$f$  &    0     &    0.13383(12)     & 0.56559(12)& 0.0061(4)\\
	O1  &  4$c$  &    0     &    0.78032(53)     &     1/4  &  0.01 \\
	O2  &  8$f$  &    0     &    0.04396(39)    & 0.11193(30) &   0.01 \\
	O3  &  8$f$  &    0     &    0.31898(32)     & 0.05842(32) &    0.01  \\ \hline
	
\end{tabular}
\end{table*}

\renewcommand{\thefigure}{S5}
\begin{figure}[t]
\begin{center}
	\includegraphics[width=75mm,trim=0cm 0cm 0cm 0cm,clip=true]{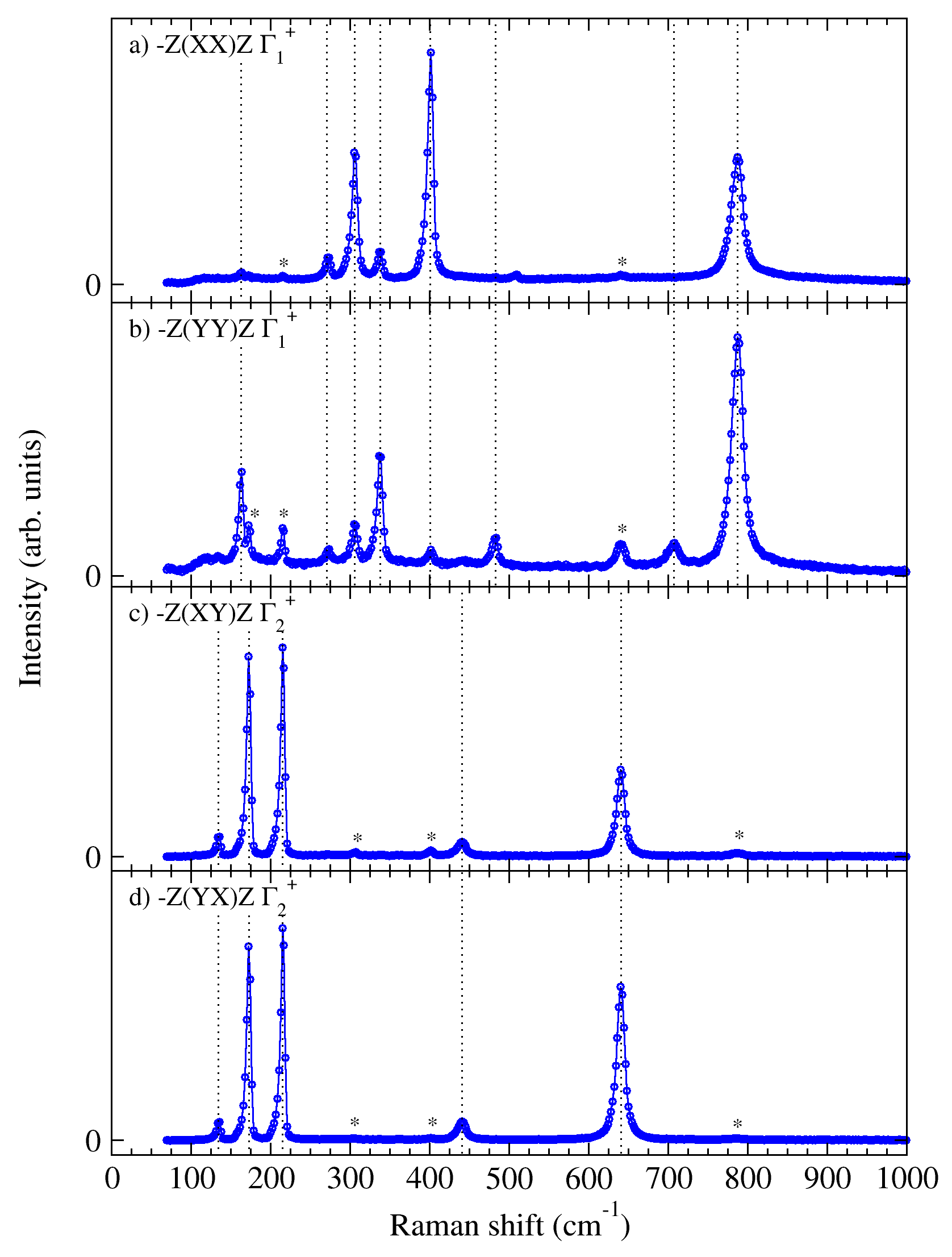}
\end{center}
\caption{Raman data for all polarizations with the different symmetry labels listed.  Weak feedthrough peaks are denoted by an astrix $*$.} 
\label{fig:Raman}
\end{figure}

\renewcommand{\thefigure}{S6}
\begin{figure}[t]
\begin{center}
	\includegraphics[width=100mm,trim=4cm 0.75cm 0cm 0.0cm,clip=true]{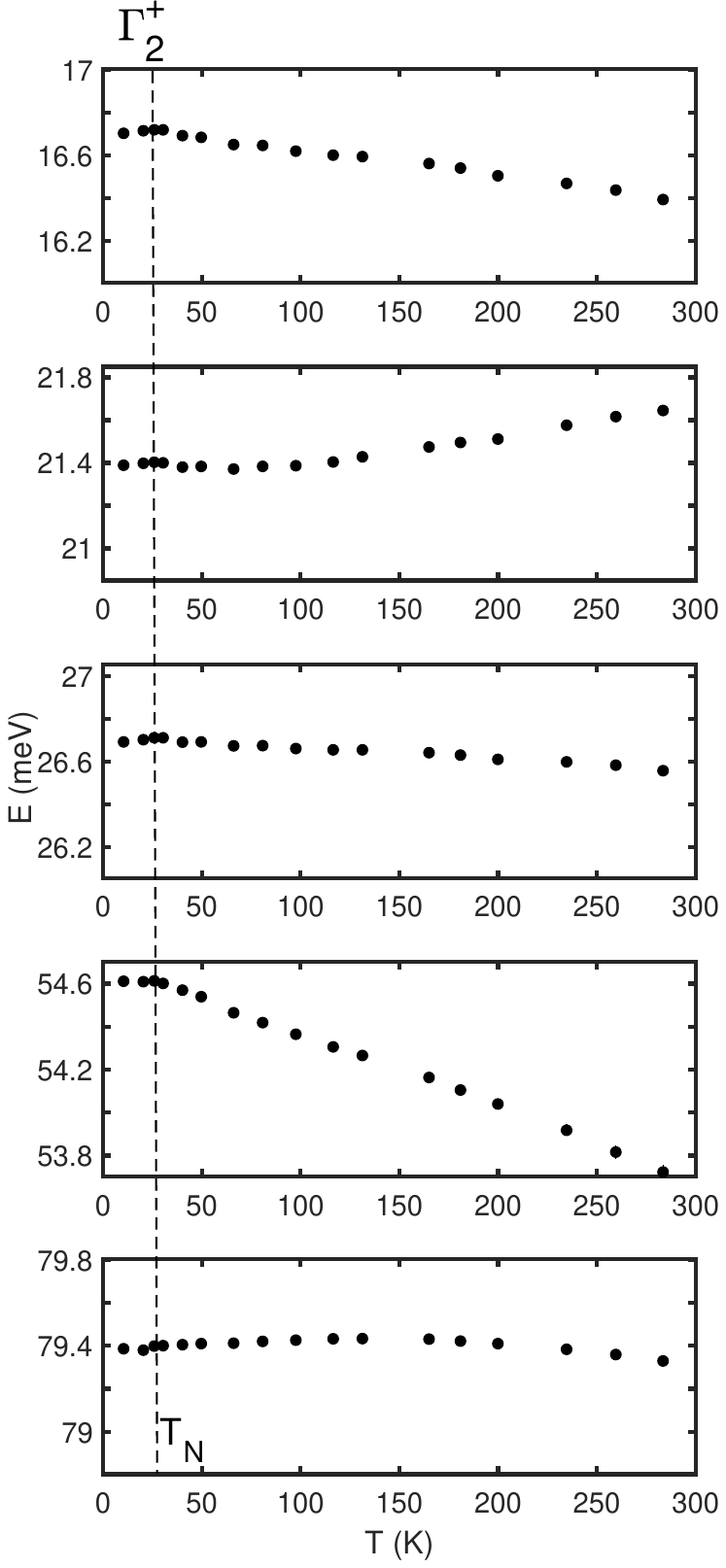}
\end{center}
\caption{The temperature dependence of the 5 $\Gamma_{2}^{+}$ modes.  A saturation in the temperature dependence of the energy values is observable at $T_{\rm N}$ with mode at $\sim$ 21 meV displaying an observable softening with temperature.} 
\label{fig:Raman_temp}
\end{figure}

\renewcommand{\thefigure}{S7}
\begin{figure}[t]
\begin{center}
	\includegraphics[width=75mm,trim=0cm 0cm 0cm 0cm,clip=true]{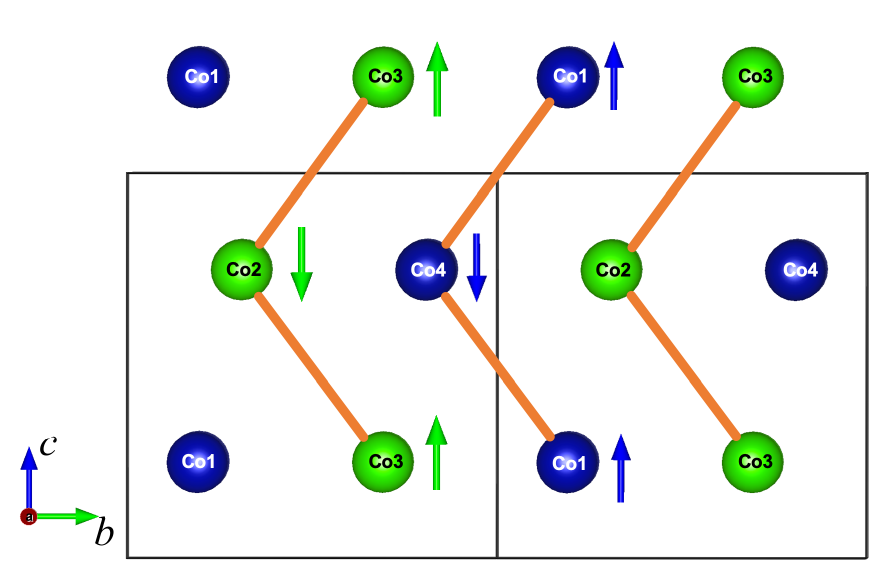}
\end{center}
\caption{A schematic representation of two distinct magnetic Co$^{2+}$ ions per unit cell. The four symmetry-equivalent crystallographic sites, defined with respect to the $Cmcm$ unit
	cell, comprising the full cobalt sublattice: Co1:[0,\textit{y},$\frac{1}{4}$];Co2:[$\frac{1}{2}$,$\frac{1}{2}$$-$\textit{y},$\frac{3}{4}$];Co3:[$\frac{1}{2}$,$\frac{1}{2}$+\textit{y},$\frac{1}{4}$];Co4:[0,1$-$\textit{y},$\frac{3}{4}$], the value of \textit{y}= 0.1911(6). The (Co1,Co4) and (Co2,Co3) moments are antiferromagnetically related to each other\cite{Kirschner19:99}.} 
\label{fig:twosites}
\end{figure}

We now discuss the point of spin-orbit coupling in this compound and the possibility of an orbital degeneracy.  Co$^{2+}$ in an octahedral crystalline electric field environment is expected to have an orbital degeneracy with $l=1$ and $S={3\over 2}$.  In such a case, the orbitally drive Jahn-Teller effect would be expected to be present with the crystal structure distorted to break the orbital degeneracy.  This is different from the spin Jahn-Teller which predicts a structural distortion breaking a ground state spin degeneracy, not an orbital degeneracy.  However, this is not expected to be the case in CoTi$_{2}$O$_{5}$ with the local crystalline electric field environment strongly distorted away from perfect octahedra.  This is confirmed in Fig. \ref{fig:spinorbit} where we plot data taken from MAPS over a broad range in energy transfer. Fig. \ref{fig:spinorbit} $(a)$ illustrates a $Q-E$ slice with data being integrated over all K and L directions given the magnetic excitations do not substantially disperse in these two directions. Fig. \ref{fig:spinorbit} $(b)$ plots an integral of the data Fig. \ref{fig:spinorbit} $(a)$ over the displayed momentum range.  If an orbital degeneracy existed, we would expect, based on previous compounds, to see a strong excitation in the range of 20-30 meV.  For example, such an excitation is clearly visible in, for example, $\alpha,\gamma$-CoV$_2$O$_6$ and CoV$_3$O$_8$ and can be assigned as a spin-orbit excitations from the $j_{eff}={1\over 2}$ to $j_{eff}={3\over 2}$ manifold.  This spin-orbit excitation is stronger than the low-energy spin-waves in this series of materials.  Such a strong magnetic excitation is not observable in Figs. \ref{fig:spinorbit} $(a,b)$ with the weak energy broadened excitations seen at $\sim$ 20 meV and at higher energies of $\sim$ 40 meV having an intensity which increases with momentum transfer indicative of a phonon origin rather than magnetic.  The lack of a clear spin-orbit excitation supports the lack of a Co$^{2+}$ orbital degeneracy in CoTi$_{2}$O$_{5}$ and the presence of spin-only magnetism.  

\section{Density Functional Theory and Raman active modes}

To corroborate the Raman data measuring optic phonons, we carried out phonon calculations only at the $\Gamma$-point using density-functional-perturbation theory (DFPT) implemented in the VASP package.  The crystal structure was first relaxed (all lattice parameters, shape, and positions of atoms) until the forces were smaller than 3 meV$/$\AA.  The cutoff energy for the plane-wave basis was set to 400 eV.  The convergence criterion for the electronic self-consistency was 10$^{-7}$ eV.  The Brillouin zone integration was carried out over 10×4×4 Monkhorst-Pack mesh. Strong Coulomb correlations were taken into account via a rotationally invariant DFT+U approach after Dudarev \textit{et al.}~\cite{Dudarev98:57, Liechtenstein95:52} The interaction parameters for Co and Ti were taken to be: $(U-J_H)_{\rm{Co}}$ = 6 eV, $(U-J_H)_{\rm{Ti}}$  = 3.5 eV.~\cite{Maksimov22:106,Streltsov05:71}  The results of this calculation are tabulated in Table \ref{tab1} and can be compared against the Raman data in Fig. \ref{fig:Raman} plotted for all polarizations illustrating the good agreement between calculations and experiments.  A summary of the symmetry notation convention used throughout the text and also linking with the notation used in Ref. \onlinecite{Cowley76:13} is given in Table \ref{tab2}.  We provide the temperature dependence of all 5 $\Gamma_{2}^{+}$ modes in Fig. \ref{fig:Raman_temp} and note the softening of the $\sim$ 21.4 meV mode in comparison to the temperature dependence of all other modes.  This is discussed in the main text of the paper.

\section{X-ray Powder Diffraction}

The polycrystalline samples of CoTi$_2$O$_5$ were synthesized using a solid-state reaction \cite{anderson2019}. The starting precursors of Co$_3$O$_4$ (Alfa Aesar, 99.95\,\%), and TiO$_2$ (Acros Chemicals, 98.98\,\%) were taken in stoichiometric ratio and ground well in an agate mortar using a pestle. Mixed powders were calcined at 900$^{\circ}$ C for 24 h. After the calcination, the powders were repeatedly reground, pelletized, and sintered for 48 hrs at 1200$^{\circ}$ C followed by 1300$^{\circ}$C for 6 h, and then quenched to room temperature. 
The single crystal used for the neutron experiments is from the same batch used by Ref. \onlinecite{Kirschner19:99}. This single crystal was grown in a four-mirror optical floating zone furnace (Crystal Systems, Inc.) in argon/oxygen mixed gas (90:10 ratio) atmosphere with a growth rate of 2–3 mm/h.

The phase purity of polycrystalline powder was confirmed using X-ray diffraction measurements at room temperature using a Rigaku SmartLab X-ray Diffractometer. The high-quality XRD data as a function of temperature were collected using a Bragg-Brentano geometry with a monochromatized Cu $K\alpha_{1}$ source, presented in this paper. A {\sc jana}2006 software~\cite{Petricek14:229} and TOPAS~\cite{TOPAS} were used for Rietveld refinements to determine the crystal structures using the orthorhombic space group $Cmcm$ which results in two unique magnetic sites (Fig. \ref{fig:twosites}). Note that, sites are assumed to be fully occupied, and Co$^{2+}$/Ti$^{4+}$ sites mixing is not included in the refinements as the difference in the scattering amplitude of these two elements is too small to be able to be distinguished by X-rays.

X-ray powder diffraction from a piece of the single crystal was performed, and the data were compared to the powder presented in Figure 1 of the main text.  Both powder patterns and refinements were identical except for a 2.5 \% impurity in the polycrystalline sample.  While our polycrystalline sample shows a $T_{\rm N}$ $\sim$ 23 K, we note that in the literature~\cite{Xu24:33,Kirschner19:99} there has been a reported variation of $T_{\rm N}$ of several Kelvin.